\documentclass{aa}
\usepackage{natbib}
\bibpunct{(}{)}{;}{a}{}{,}
\usepackage{graphicx}
\usepackage{txfonts}

\def\0{\hspace*{0.5em}}
\newcommand{\kms}{km~s$^{-1}$}

\newcommand{\wray}{\object{Wray~977}}
\begin{document}

  \title{VLT/UVES spectroscopy of Wray 977, the hypergiant companion to
  the X-ray pulsar \object{GX301$-$2}\thanks{Based on observations obtained at the
  European Southern Observatory (57.D-0410,68.D-0568(A)) and with the Infared
  Space Observatory (ISO), a project of the European Space Agency
  (ESA) with the participation of ISAS and NASA.}}

\author{L.\ Kaper\inst{1} \and A. van der Meer\inst{1} \and F.\
  Najarro\inst{2}} 

   \offprints{L. Kaper}

   \institute{Astronomical Institute ``Anton Pannekoek'', University of
  Amsterdam, Kruislaan 403, 1098 SJ Amsterdam, The Netherlands\\
  \email{lexk@science.uva.nl; ameer@science.uva.nl} \and
  Instituto de Estructura de la Materia, CSIC, Serrano 121, 29006
  Madrid, Spain\\ \email{najarro@damir.iem.csic.es} }

   \date{Received; Accepted}

   \authorrunning{L.\ Kaper, A. van der Meer \& F. Najarro}
   \titlerunning{Wray 977, the hypergiant companion to GX~301-2}

\abstract{Model atmosphere fits to high-resolution optical spectra of
\wray\ confirm the B~hypergiant classification of the massive
companion to the X-ray pulsar \object{GX301$-$2}. The models give a
radius of 62~$R_{\sun}$, an effective temperature of 18,100~K and a
luminosity of $5 \times 10^{5} \, L_{\sun}$. These values are somewhat
reduced compared to the stellar parameters of \wray\ measured
previously. The deduced mass-loss rate and terminal velocity of the
stellar wind are $10^{-5}$~M$_{\sun}$~yr$^{-1}$ and 305~\kms,
respectively. The interstellar $\ion{Na}{i}$ D absorption indicates
that \wray\ is located behind the first intersection with the
Sagittarius-Carina spiral arm (1-2.5~kpc) and probably belongs to the
stellar population of the Norma spiral arm at a distance of
$3-4$~kpc. The luminosity derived from the model atmosphere is
consistent with this distance ($3$~kpc). The luminosity of the
wind-fed X-ray pulsar ($L_{X} \sim 10^{37}$~erg~s$^{-1}$) is in good
accordance with the Bondi-Hoyle mass accretion rate. The spectra
obtained with UVES on the {\it Very Large Telescope} (VLT) cover a
full orbit of the system, including periastron passage, from which we
derive the radial-velocity curve of the B hypergiant. The measured
radial-velocity amplitude is $10 \pm 3$~\kms\ yielding a mass ratio $q
= M_{X}/M_{\rm opt} = 0.046 \pm 0.014$. The absence of an X-ray
eclipse results in a lower limit to the mass of \wray\ of
39~M$_{\sun}$. An upper limit of 68 or 53~M$_{\sun}$ is derived for
the mass of \wray\ adopting a maximum neutron star mass of 3.2 or
2.5~M$_{\sun}$, respectively. The corresponding lower limit to the
system inclination is $i > 44^{\circ}$, supporting the view that the
dip in the X-ray lightcurve is due to absorption by the dense stellar
wind of \wray\ \citep{Leahy02}.  The ``spectroscopic'' mass of \wray\
is $43 \pm 10$~M$_{\sun}$, consistent with the range in mass derived
from the binarity constraints. The mass of the neutron star is $1.85
\pm 0.6$~M$_{\sun}$.  Time series of spectral lines formed in the
dense stellar wind (e.g. $\ion{He}{i}$ 5876~\AA\ and $\rm H \alpha$)
indicate the presence of a gas stream trailing the neutron star in its
orbit. The long-term behaviour of the $\rm H \alpha$ equivalent width
exhibits strong variations in wind strength; the sampling of the data
is insufficient to conclude whether a relation exists between wind
mass-loss rate and pulsar spin period.

   \keywords{Stars: binaries: close -- Stars: evolution -- Stars:
individual: Wray 977 -- Stars: pulsars: individual: \object{GX301$-$2} -- Stars:
supergiants -- X-rays: stars}

}

  \maketitle

%

\section{Introduction}

\wray\ (BP~Cru) is the B-supergiant companion to the X-ray pulsar
\object{GX301$-$2}. Comparison of the mass functions derived for
high-mass X-ray binaries (HMXBs) harbouring an X-ray pulsar shows that
\wray\ is the most massive OB-star companion in these systems
\citep{Nagase89,Bildsten97}; HD153919, the O6.5 Iaf$^{+}$ companion to
4U1700-37, may have a higher mass ($M_{\rm opt} = 58 \pm 11$,
\citealt{Clark02}), but 4U1700-37 has not been proven to be an X-ray
pulsar \citep[e.g.][]{Hon04}. HMXBs are divided into two sub-groups:
the Be/X-ray binaries (mainly X-ray transients) and OB-supergiant
systems (of which \wray\ is a member). The transient character of the
Be/X-ray binaries relates to ``outburst'' phases of the Be-type
companion, which occur at irregular intervals, separated by years to
decades. During such an outburst phase the Be star ejects matter which
forms an equatorial disc. The crossing of the compact companion
through the Be-star's equatorial disc leads to X-ray outbursts which
recur with the orbital period. In OB-supergiant systems the X-ray
source accretes from the strong stellar wind or is fed by Roche-lobe
overflow \citep[see][]{Kaper01}. In the latter case the much higher
accretion rate results in an about 100 times higher X-ray luminosity
($\sim 10^{38}$~erg~s$^{-1}$) and a short X-ray pulse period (seconds
rather than minutes) in comparison to wind-fed systems.

The mass function derived from pulse-timing analysis indicates that
the mass of \wray\ is larger than 31.8~M$_{\sun}$
\citep{Sato86,Koh97}. Knowledge of the mass of \wray\ is important, because
this information is used to determine the empirical lower mass limit
for black-hole formation in a massive binary
\citep{VandenHeuvel84,Ergma98,Wellstein99}. The progenitor of \object{GX301$-$2} was
originally the most massive star in the system and left a neutron star
(the X-ray pulsar) after the supernova explosion. The observed absence
of X-ray eclipses sets an upper limit to the system's inclination,
given an estimate of the radius of \wray. The larger the radius, the
lower the inclination, and the higher the lower limit to its (present)
mass. Furthermore, the high mass of \wray\ may result in the formation
of a black-hole -- neutron-star binary when \wray\ ends its life as a
supernova or a gamma-ray burst.

Based on a reclassification of its optical spectrum,
\citet{Kaper95} proposed that \wray\ is a B1~Ia+ hypergiant
rather than a normal B supergiant \citep{Parkes80}. Since a hypergiant
is larger and more luminous than a supergiant of the same spectral
type, this results in a larger mass (48~M$_{\sun}$) and larger distance
(5.3~kpc) of \wray\ than thought before. In fact, the radius adopted
in \citet{Kaper95} ($R_{\star} = 87 \, R_{\sun}$) is larger than the Roche
(and tidal) radius at periastron passage of this very eccentric
($e=0.462$) system. The interpretation of the pulse-period history
(spin-up) of \object{GX301$-$2} over the last decade \citep{Pravdo95} and a
reconsideration of the parameters of the binary system \citep{Koh97}
suggest that the distance, stellar radius and thereby the mass of
\wray\ are less than proposed in \citet{Kaper95}.

From the cyclical occurence of X-ray flares, \citet{Watson82}
determined the orbital period of the system ($P_{\rm orb} =
41.5$~d). It turns out that the periodic flare occurs just {\it
before} periastron passage \citep[$\sim 2$d,
\,][]{White84,Sato86,Chichkov95}. A similar X-ray lightcurve has been
observed for 4U~1907+09 \citep{IntZand98}; in this system the massive
companion star also is a luminous supergiant with a dense stellar wind
\citep[O8-O9 Ia,][]{Cox05a}.  Calculations by \citet{Stevens88} of the
dynamical effects of the neutron star on the stellar wind showed that
a highly increased mass-loss rate from the primary can be expected at
periastron passage. This provided the physical basis for the
suggestion by \citet{Haberl91} and \citet{Leahy91} that the observed
pre-periastron X-ray flares are due to the enhanced accretion rate
during the passage of the neutron star through a gas stream in the
stellar wind.  \citet{Pravdo95} found that also near apastron passage
a periodic X-ray flare occurs; the gas-stream model cannot explain the
apastron flare very well. They propose that both the asymmetry of the
pre-periastron flare and the presence of the near-apastron flare can
be explained by an equatorially enhanced stellar wind or a
circumstellar disc around \wray. The X-ray flares would occur when the
X-ray source moves through the disc which is slightly inclined with
respect to the orbital plane of the X-ray pulsar, as in the case of
Be/X-ray binaries.


The rapid spin-up episodes of \object{GX301$-$2} discovered by \citet{Koh97}
suggest the formation of temporary accretion discs. The long-term
spin-up trend of the X-ray pulsar observed since 1984 might be
entirely due to such brief spin-up episodes. Numerical simulations
carried out by \citet{Layton98} confirm that tidal stripping in
eccentric-orbit X-ray binaries can produce periodic flares. However,
the tidally stripped mass only accretes when the supergiant is close
to corotation at periastron, and the resulting X-ray flare occurs well
{\it after} periastron passage ($\phi\sim 0.2$) and not at
apastron. The calculations further show that a transient accretion
disc forms when the neutron star accretes from the tidal stream and
persists for roughly half the binary period. This produces an extended
epoch (many days) of spin-up reminiscent of the spin-up episodes
observed by \citet{Koh97}. \citet{Layton98} remark that the X-ray
flares are more likely to be due to enhanced accretion from an
equatorial disc than to tidal stripping at periastion.
\begin{table}[!t]
\caption[]{Optical and near-infrared broad-band photometric parameters
          of \wray\ collected from literature; the infrared fluxes
          were observed with ISO.  The fluxes are dereddened using the
          extinction law of \citet{Mathis90} and $E(B-V)=1.9$. (1)
          \citet{Bord76}; (2) \citet{Hammerschlag76}; (3)
          \citet{VanDishoeck89}; (4) \citet{Coe97}; (5)
          \citet{Glass79}; (6) ISO (see
          Sect.~\ref{iso_observations}).}
\begin{center}
\begin{tabular}{llllll}
\hline \hline
Passband & $\lambda$ & Mag & $f_{\nu}$ & $A_{\lambda}$ (mag) & $f_{\nu}$ (cor) \\
         & ($\mu$m)  &     & (Jy)      & ($R_{V}=3.1$)       & (Jy)  \\
 \hline
   U (1)     & \00.365      & 13.01    & 0.012  &  9.18         & 55.23 \\
   B (1)     & \00.44       & 12.59    & 0.041  &  7.80         & 53.87 \\
   V (2)     & \00.55       & 10.83    & 0.177  &  5.89         & 40.26 \\
   I (3)     & \00.90       & \07.6    & 2.049  &  2.82         & 27.52 \\
   J (4)     & \01.25       & \06.83   & 2.817  &  1.66         & 13.00 \\
   H (4)     & \01.65       & \06.11   & 3.526  &  1.04         & \09.19  \\
   K (4)     & \02.2        & \05.72   & 3.194  &  0.63         & \05.71  \\
   L (5)     & \03.4        & \05.25   & 2.304  &  0.30         & \03.04  \\
   LW2 (6)   & \06.75       &          & 0.837  &  0.12         & \00.94  \\     
   LW10 (6)  & 11.5         &          & 0.517  &  0.16         & \00.60  \\
   PHT03 (6) & 25.0         &          & 0.089  &  0.08         & \00.10  \\
\hline
\end{tabular}
\end{center}
\label{tabphot}
\end{table}

The main motivation of this paper is to provide a better and
quantitative estimate of the stellar parameters of \wray\ based on
model atmosphere fits to the optical line spectrum and the energy
distribution. The mass ratio of the system is determined by measuring
the radial-velocity orbit of \wray. Further, we monitored \wray\ to
search for the presence of a gas stream in the system. In the next
section we describe the observations.  Constraints on the distance and
mass of \wray\ are discussed in Sect.~\ref{constraints}. In
Sect.~\ref{modelling} we present the results of the spectrum
modelling. In Sect.~\ref{interaction} we study the interaction between
the X-ray pulsar and the stellar wind, search for the presence of a
gas stream and show that the observed X-ray flux is consistent with
that expected from Bondi-Hoyle accretion.  In the last section we
summarize the conclusions and discuss the implications of the derived
stellar parameters for the nature and evolutionary status of this
binary system.

\section{Observations}
\label{observations}

The results presented in this paper are based on a variety of archival
and recently obtained optical and infrared observations. Broad-band
photometry is needed to construct the energy distribution of
\wray. The {\it Infrared Space Observatory} (ISO) has detected \wray\
at infrared wavelengths. The bound-free and free-free emission
produced by the stellar wind in this wavelength region sets a
constraint on the wind mass-loss rate, which is an important parameter
in modelling the spectrum of \wray. High-resolution optical spectra
are used to derive model-atmosphere parameters, to measure the
radial-velocity curve and to search for the presence of a gas stream
in the system.
\begin{figure*}[!t]
\begin{center}
\includegraphics[height=12cm,angle=90]{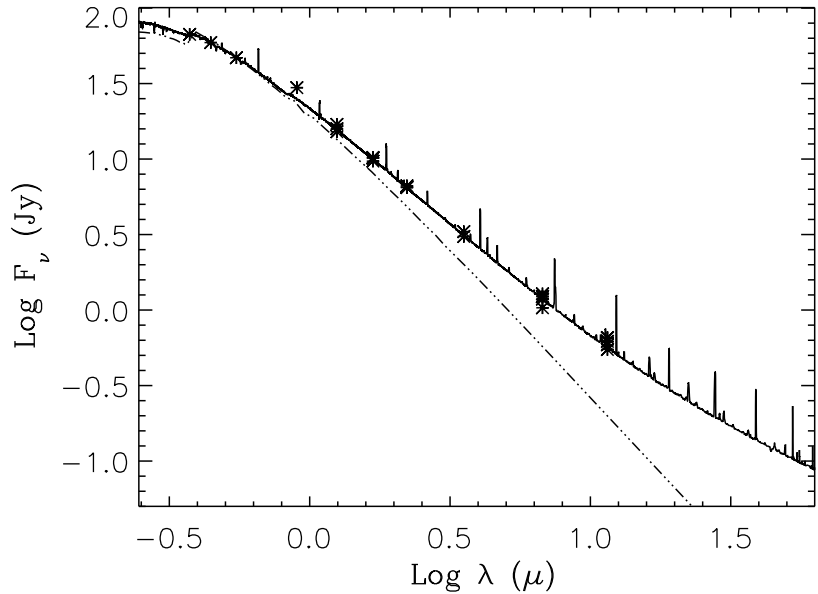}
\caption[]{Deredenned energy distribution of Wray 977. The current
          best model (solid; Sect.~\ref{modelling}) is shown together with a Kurucz model
          with the same temperature ($T_{\rm eff} = 18\,100$~K,
          dashed-dotted).  The observations (asterisks) have been
          deredenned assuming $R_{\rm V} = 3.1$ and $E(B-V) = 1.96$, which
          results in a distance of 3.04~kpc for our model with
          $R_{2/3}=70\,R_{\sun}$.}
\label{fig_energydist}
\end{center}
\end{figure*} 

\subsection{Optical and near-infrared photometry}

Table~\ref{tabphot} lists the broad-band optical and near-infrared
photometric parameters collected from literature. The resulting energy
distribution of \wray\ ($V=10.8$) is displayed in
Fig.~\ref{fig_energydist}. For comparison, a Kurucz model with the same
$T_{\rm eff}= 18,100$~K, $\log{g}=3.0$ and $E(B-V)=1.96$ (assuming
$R_{V}=3.1$) is shown. A strong infrared excess with respect to the Kurucz
model is apparent, caused by the free-free emission produced by the
dense stellar wind.

The line of sight towards \wray\ passes the edge of the Southern
Coalsack, one of the most prominent dark nebulae in the southern Milky
Way \citep[e.g.,][]{Nyman89} and continues through the
Sagittarius-Carina spiral arm. The strong interstellar extinction
makes the blue hypergiant appear red in the telescope. \citet{Kaper95}
adopted a colour excess of $E(B-V)=1.9$, based on the value listed
in \citet{VanDishoeck89}. \citet{VanGenderen96} derive $E(B-V)=1.65
\pm 0.10$ based on Walraven photometry and $E(B-V)=2.0$ when fitting
the energy distribution with a black-body with a temperature of
$21\,100$~K. The same value for the colour excess is obtained from
Str\"{o}mgren photometry. \citet{VanGenderen96} have no explanation
for these inconsistencies, but prefer the lower value of 1.65. In this
paper we adopt $E(B-V)=1.96$, based on a comparison with model
atmospheres (see Sect.~\ref{modelling}).

\subsection{ISO observations}
\label{iso_observations}

\wray\ is one of the targets in our ISO program on HMXBs to study the
disrupted stellar wind and the accretion flow towards the X-ray source
in these systems \citep{Kaper97}. The infrared wavelength domain has
proven to provide a valuable diagnostic in studying the stellar-wind
structure of early-type stars \citep[see][]{Lamers84}. Free-free and
free-bound emission is produced in the inner part of the stellar wind
where most of the acceleration takes place and can be modelled in a
quite straightforward way.

ISO observations of \wray\ were performed with ISOCAM's broad-band
filters LW2 and LW10 (centered at 6.75 and 11.5 $\mu$m) and with
ISOPHOT (25 $\mu$m), see Table~\ref{tabphot}. The observations were
obtained (at the three different wavelengths) in August 1996 and in
July and August 1997, covering four orbital phases of the system. We
used the ISOCAM (CIA) and ISOPHOT Interactive Analysis (PIA) packages and
the supplied auto-analysis products to reduce the data. In
Table~\ref{tabphot} we give the (orbital) average of the
background-subtracted flux values. The orbital variation in flux is at
the 5-10~\% level.
\begin{table}[!t]
\caption[]{Log of observations of high-resolution spectra obtained
          with the CAT/CES. Column (1) lists the heliocentric Modified
          Julian Date; (2) the orbital phase according to the combined
          ephemeris of \citet{Sato86} and \citet{Koh97}, see
          Table~\ref{taborbit}; (3) the corresponding true anomaly
          $\nu$; (4) the central wavelength ($\lambda_{\rm c}$) and
          (5) the exposure time. The radial velocity of the spectrum
          (6) is measured in the heliocentric frame, the listed error
          corresponds to the standard deviation of the mean velocity
          of all lines measured in a given spectrum.}
\begin{center}
\begin{footnotesize}
\begin{tabular}{llllll}
\hline \hline
MJD & Phase & $\nu$ & $\lambda_{\rm c}$ & $t_{\rm exp}$ & $v_{\rm rad}$ \\
   &        & (deg) & (\AA)         & (s)           & (\kms)        \\ 
\hline
\multicolumn{6}{c}{\textit{CAT/CES May 1996}} \\
\hline
50213.9917 & 0.0065 &     & 6562 & 1800 & \\
50214.0174 & 0.0071 &     & 6562 & 1800 & \\
50214.0535 & 0.0080 &     & 5876 & 2700 & \\
50214.1417 & 0.0101 & 4   & 5695 & 2700 & $-21.4 \pm 6.8$ \\
50214.2035 & 0.0116 &     & 6562 & 2700 & \\
50214.2590 & 0.0129 &     & 5876 & 2700 & \\
50214.9820 & 0.0304 &     & 6562 & 2700 & \\
50215.0222 & 0.0313 &     & 5876 & 2700 & \\
50215.0924 & 0.0330 &     & 6562 & 2700 & \\
50215.1181 & 0.0336 & 35  & 5695 & 2700 & $-23.0 \pm 4.7$ \\
50215.2632 & 0.0371 &     & 5876 & 2700 & \\
50215.9833 & 0.0545 &     & 6562 & 2700 & \\
50216.0201 & 0.0554 &     & 5876 & 2700 & \\
50216.1000 & 0.0573 & 56  & 5695 & 2700 & $-27.8 \pm 4.0$ \\
50216.1375 & 0.0582 &     & 6562 & 2700 & \\
50216.1931 & 0.0595 &     & 5876 & 2700 & \\
50216.2472 & 0.0609 &     & 6562 & 2700 & \\
50216.9903 & 0.0788 &     & 6562 & 2700 & \\
50217.0396 & 0.0799 &     & 5876 & 2700 & \\
50217.1285 & 0.0821 & 74  & 5695 & 2700 & $-30.1 \pm 5.0$ \\
50217.1680 & 0.0830 &     & 6562 & 2700 & \\
50217.2215 & 0.0843 &     & 5876 & 2700 & \\
50217.9791 & 0.1026 &     & 6562 & 2700 & \\
50218.0222 & 0.1036 &     & 5876 & 2700 & \\
50218.1028 & 0.1056 & 88  & 5695 & 2700 & $-31.5 \pm 3.7$ \\
50218.1423 & 0.1065 &     & 6562 & 2700 & \\
50218.1986 & 0.1079 &     & 5876 & 2700 & \\
50218.9812 & 0.1267 &     & 6562 & 2700 & \\
50219.0187 & 0.1276 &     & 5876 & 2700 & \\
50219.1000 & 0.1296 &     & 6562 & 2700 & \\ 
50219.1014 & 0.1296 & 100 & 5695 & 2700 & $-34.7 \pm 4.1$ \\
50219.2041 & 0.1321 &     & 5876 & 2700 & \\
\hline
\end{tabular}
\end{footnotesize}
\end{center}
\label{tablogces}
\end{table}
\begin{table}[!t]
\caption[]{Log of observations of high-resolution spectra obtained
          with the VLT/UVES. Column (1) lists the heliocentric
          Modified Julian Date; (2) the orbital phase according to the
          combined ephemeris of \citet{Sato86} and \citet{Koh97}, see
          Table~\ref{taborbit}; (3) the corresponding true anomaly
          $\nu$. The central wavelengths (4) corresponding to the used
          dichroic setting \citep[DIC\#1,~][]{Kaufer00} of VLT/UVES
          are 4035, 5108, and 6160~\AA. Column (5) lists the exposure
          time. The radial velocity of the spectrum (6) is measured in
          the heliocentric frame, the listed error corresponds to the
          standard deviation of the mean velocity of all lines
          measured in a given spectrum.}
\begin{center}
\begin{footnotesize}
\begin{tabular}{llllll}
\hline \hline
MJD & Phase & $\nu$ & $\lambda_{\rm c}$ & $t_{\rm exp}$ & $v_{\rm rad}$ \\
   &        & (deg) & (\AA)         & (s)           & (\kms)        \\ 
\hline
\multicolumn{6}{c}{\textit{VLT/UVES January - February 2002}} \\ 
\hline
52289.3692 & 0.0180 & \0\02 & DIC\#1 & 1120 & $-15.4 \pm 3.2$ \\
52292.2937 & 0.0885 & \078 & DIC\#1 & 1120 & $-20.7 \pm 2.8$ \\
52298.2532 & 0.2321 & 133 & DIC\#1 & 1120 & \0\0$0.9 \pm 2.6$ \\
52309.3431 & 0.4993 & 180 & DIC\#1 & 1120 & \0\0$7.7 \pm 2.2$ \\
52315.1934 & 0.6403 & 202 & DIC\#1 & 1120 & \0$-6.1 \pm 1.8$ \\
52321.1829 & 0.7846 & 231 & DIC\#1 & 1120 & \0$-0.3 \pm 2.3$ \\
52326.2084 & 0.9057 & 278 & DIC\#1 & 1120 & \0\0$7.3 \pm 2.4$ \\
52328.1649 & 0.9529 & 312 & DIC\#1 & 1120 & $-12.2 \pm 2.1$ \\
52329.2378 & 0.9787 & 337 & DIC\#1 & 1120 & $-17.4 \pm 2.0$ \\ 
52330.2073 & 1.0021 & 362 & DIC\#1 & 1120 & $-18.1 \pm 2.0$ \\ 
52332.2167 & 1.0505 & 411 & DIC\#1 & 1120 & $-12.2 \pm 1.7$ \\
\hline
\end{tabular}
\end{footnotesize}
\end{center}
\label{tabloguves}
\end{table}
\begin{figure*}[!t]
\begin{center}
\includegraphics[width=15cm]{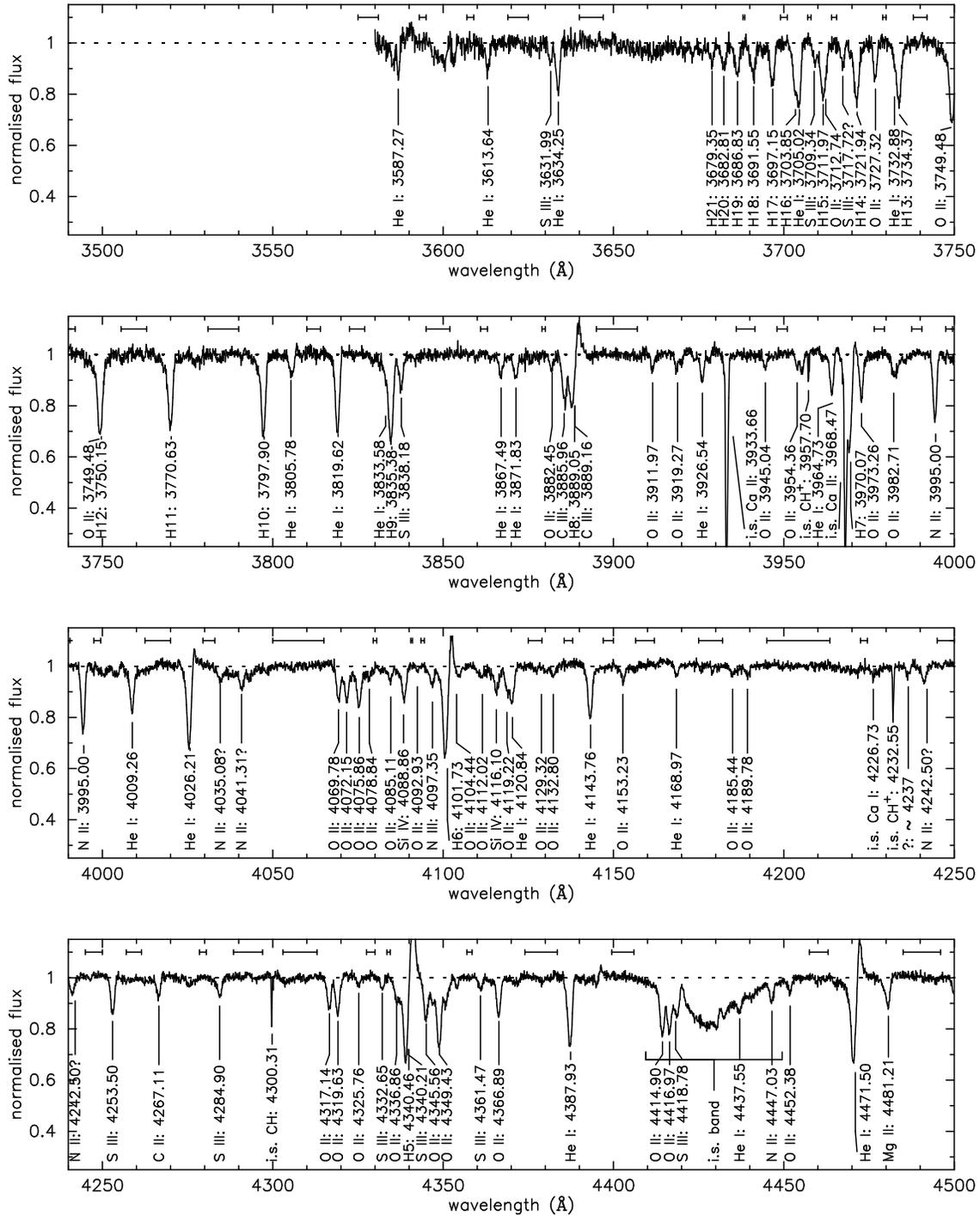}
\caption[]{The blue VLT/UVES spectrum of \wray\ ($\phi = 0.0$)
          includes the Balmer series limit at 3655~\AA. The stronger
          hydrogen and helium lines show P~Cygni-type profiles
          produced by the dense stellar wind. The large reddening
          towards \wray\ is reflected by the prominent interstellar
          lines. The regions selected to normalise the continuum are
          indicated above the spectrum.}
\label{figbluespec}
\end{center}
\end{figure*}
\begin{figure*}[!t]
\begin{center}
\includegraphics[width=15cm]{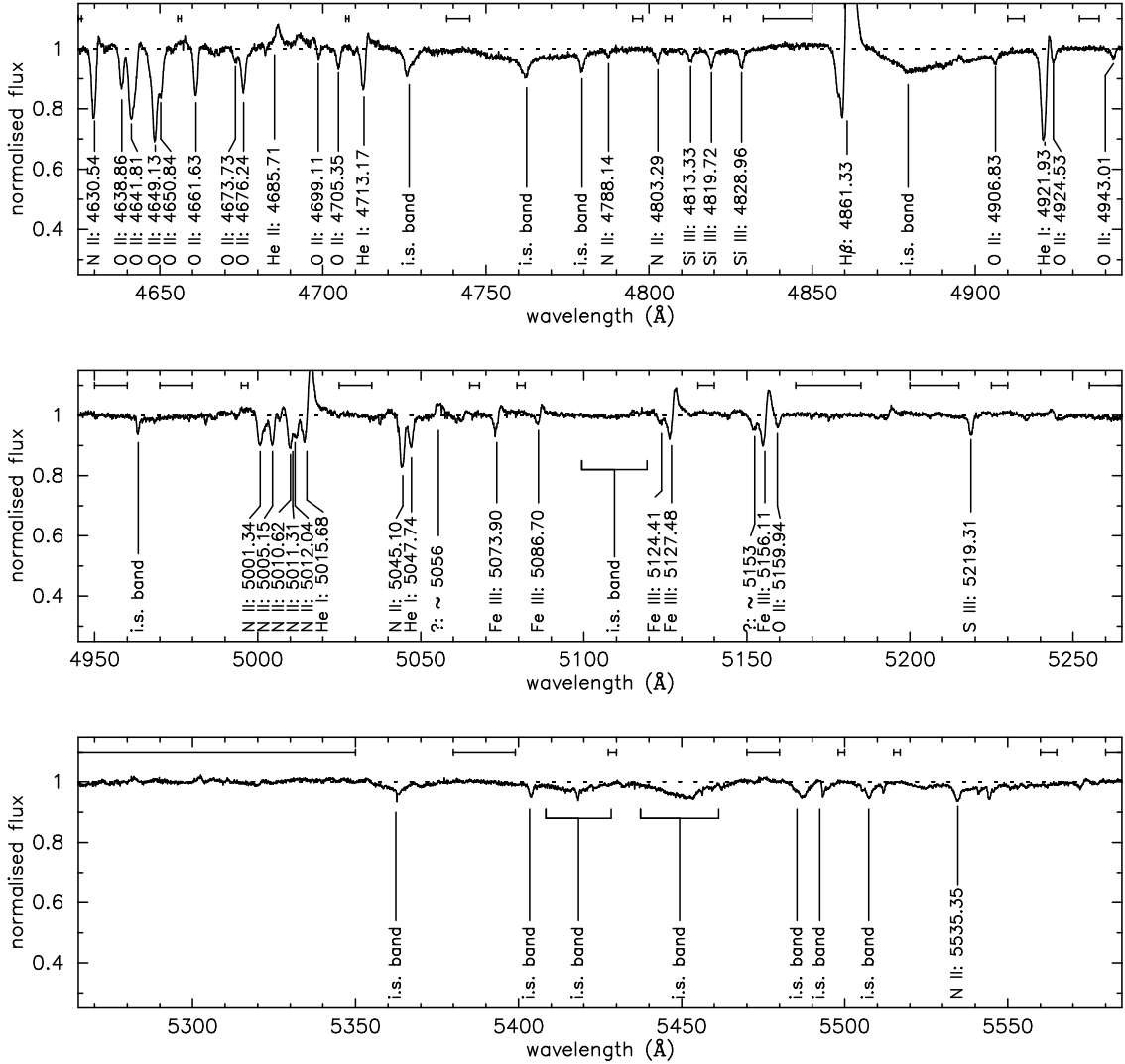}
\caption[]{The most remarkable features in the green-yellow VLT/UVES
          spectrum of \wray\ ($\phi = 0.0$) are the strong P-Cygni
          profile of $\rm H \beta$ and the $\ion{Fe}{iii}$ lines
          characteristic of a B hypergiant in the region
          5050--5150~\AA. The regions selected to normalise the
          continuum are indicated above the spectrum.}
\label{figred1spec}
\end{center}
\end{figure*}
\begin{figure*}[!t]
\begin{center}
\includegraphics[width=15cm]{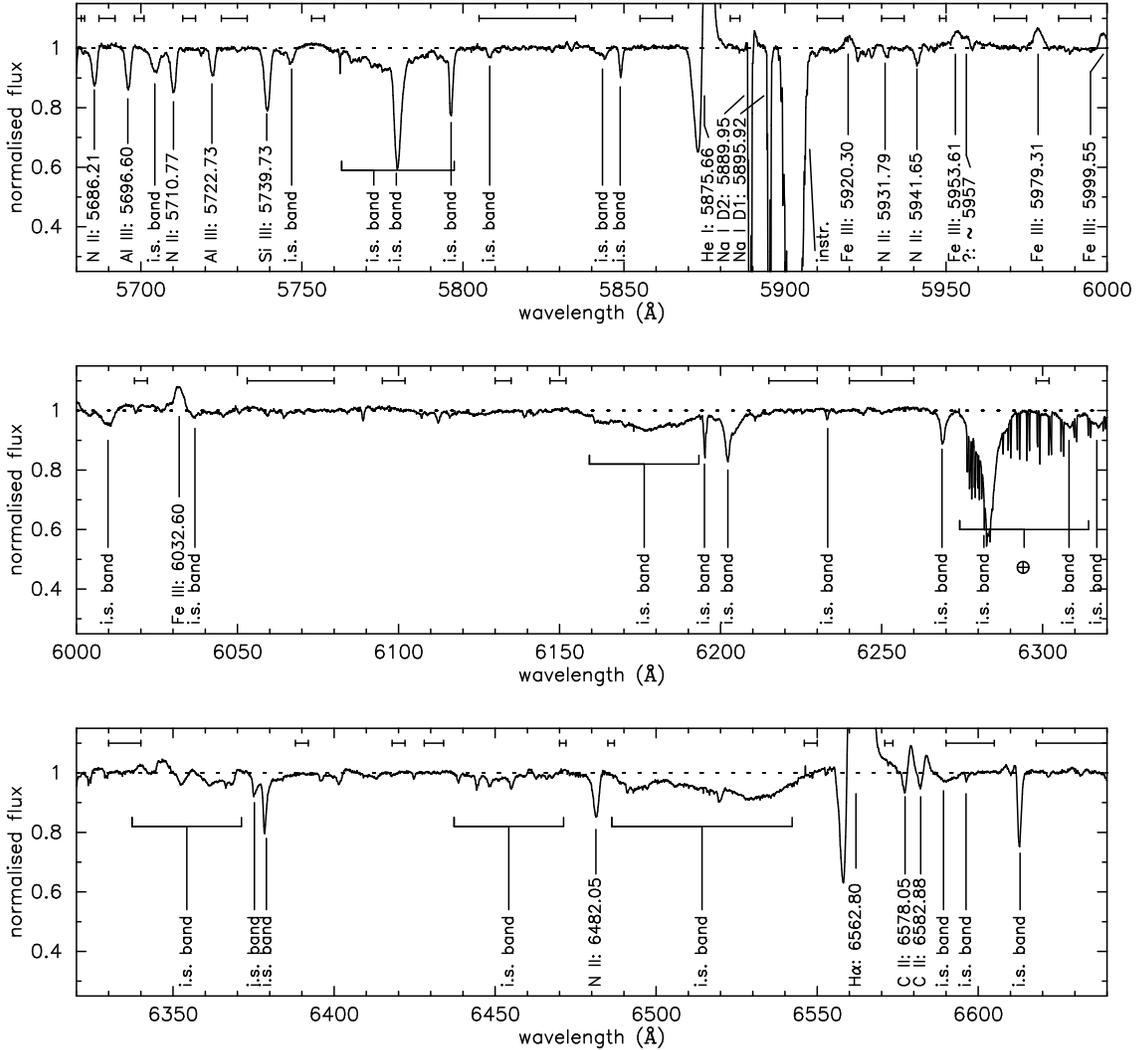}
\caption[]{The red VLT/UVES spectrum of \wray\ ($\phi = 0.0$)
          includes a very strong $\rm H \alpha$ profile (P-Cygni
          emission $\sim$3 times continuum) and $\ion{Fe}{iii}$
          emission lines around 5950~\AA. The absorption feature near
          5900~\AA\ is due to a CCD defect. The regions selected to
          normalise the continuum are indicated above the spectrum.}
\label{figred2spec}
\end{center}
\end{figure*}

\subsection{High-resolution optical spectra}

Optical spectra of \wray\ were obtained with the 1.4m {\it Coud\'{e}
Auxiliary Telescope} (CAT) and Coud\'{e} Echelle Spectrograph (CES) in
May 1996 at the European Southern Observatory in La Silla and with the
high-resolution Ultraviolet and Visual Echelle Spectrograph (UVES) on
the {\it Very Large Telescope} (VLT) in January and February 2002 at
ESO Paranal Observatory. The CAT/CES observations were obtained in
coordination with simultaneous X-ray observations with the {\it Rossi
X-ray Timing Explorer} (RXTE) satellite during periastron passage of
\object{GX301$-$2} \citep{Mukherjee04}. The VLT/UVES observations were
carried out in service mode. The log of observations of the CAT/CES
and VLT/UVES spectra is given in Table~\ref{tablogces} and
\ref{tabloguves}, respectively.

All spectra were bias subtracted, flatfield corrected, (optimally)
extracted and wavelength calibrated using the MIDAS echelle (UVES) and
long-slit (CES) data reduction packages. The resolving power of the
spectra is $R\sim 40\,000$. The signal-to-noise ratio of the CES
spectra is typically $S/N \sim 150$. The quality of the UVES spectra
improves as a function of wavelength, because of the strong
interstellar extinction towards \wray: $S/N \sim 65, \, 140, \, 240$
at 4050, 5100, 6250~\AA, respectively. The single order CES spectra
cover only $\sim 65$~\AA. The wavelength regions covered by the UVES
echelle spectra are 3570--4500~\AA, 4625--5590~\AA\ and
5675--6645~\AA. The spectra were rectified using a spline fit through
carefully selected continuum regions.

The UVES spectra are displayed in Figs.~\ref{figbluespec} -
\ref{figred2spec}; labels show the line identifications. The blue
spectrum (Fig.~\ref{figbluespec}) covers the Balmer series from
H$\gamma$ up to (at least) $\ion{H}{}21$ at 3679~\AA. The latter
observation indicates a relatively low electron density $N_{\rm e}$ and
thus a low surface gravity; at higher electron densities Stark
broadening will result in the merging of higher Balmer series lines
into a quasi-continuum. A low surface gravity is expected in case of a
super- or hypergiant. The stronger hydrogen and helium lines show a
P~Cygni-type profile, as expected for a dense stellar wind. Several
metallic lines produced by e.g. $\ion{C}{ii}$, $\ion{N}{ii}$,
$\ion{O}{ii}$, $\ion{Si}{iv}$ are detected in the blue part of the
spectrum.  The interstellar spectrum is very prominent: both atomic
interstellar lines and strong diffuse interstellar bands (DIBs) are
present \citep[see][]{Cox05b}.

The green to red spectrum (Figs.~\ref{figred1spec} -
\ref{figred2spec}) contains fewer spectral lines, as expected for an
early-type star. No $\ion{He}{ii}$ lines other than 4686~\AA\ are
detected, indicating that \wray\ has a spectral type later than
O. Note the $\ion{Fe}{iii}$ P-Cygni lines around 5100~\AA\ and the
broad $\ion{Fe}{iii}$ emission near 5950~\AA.  \citet{Wolf85} propose
that the $\ion{Fe}{iii}$ lines of multiplets 115 and 117 can be used
to discriminate between normal B supergiants and B hypergiants
(B~Ia+). These lines originate from the same upper level at an energy
of 20.87-20.89~eV above the ground level. This energy is very close to
the energy of a forbidden $\ion{He}{i}$ line from the singlet to the
triplet configuration at 20.97~eV. The resonance transition of
$\ion{He}{i}$ is a bit further away (21.22~eV, 584.3~\AA) and likely a
very strong line. \citet{Wolf85} suggest that the upper level of the
observed $\ion{Fe}{iii}$ emission lines is pumped by
$\ion{He}{i}$. With this explanation it can be understood that these
$\ion{Fe}{iii}$ lines are luminosity sensitive, since the proposed
fluorescence excitation mechanism only works under non-LTE
conditions. Thus, according to this diagnostic, \wray\ would be a B
hypergiant. The feature at 5900~\AA\ is caused by a CCD defect.
\begin{figure*}[!t]
\begin{center}
\includegraphics[angle=-90,width=10.6cm]{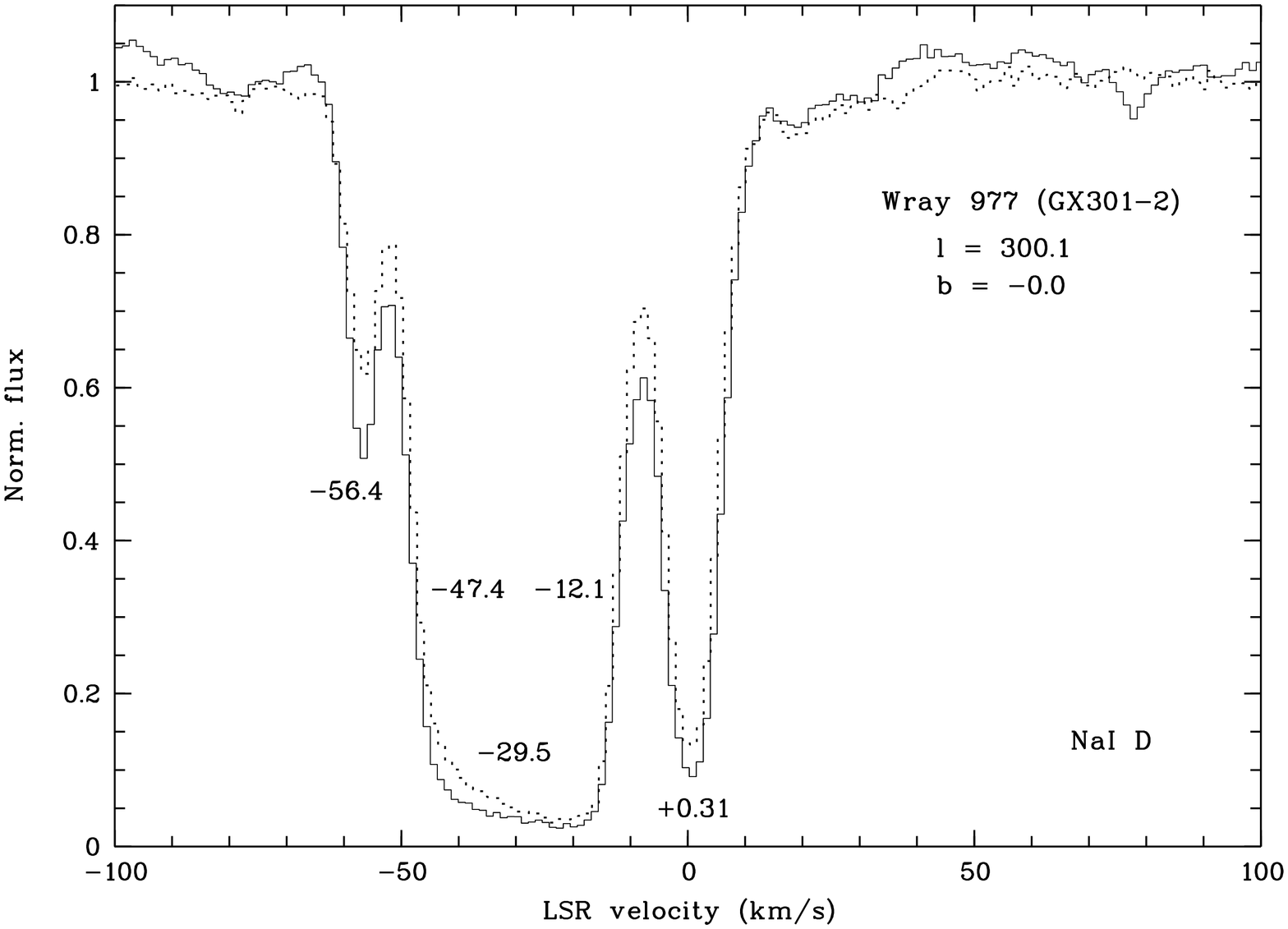}
\\[-0.2cm]
\includegraphics[angle=-90,width=10.6cm]{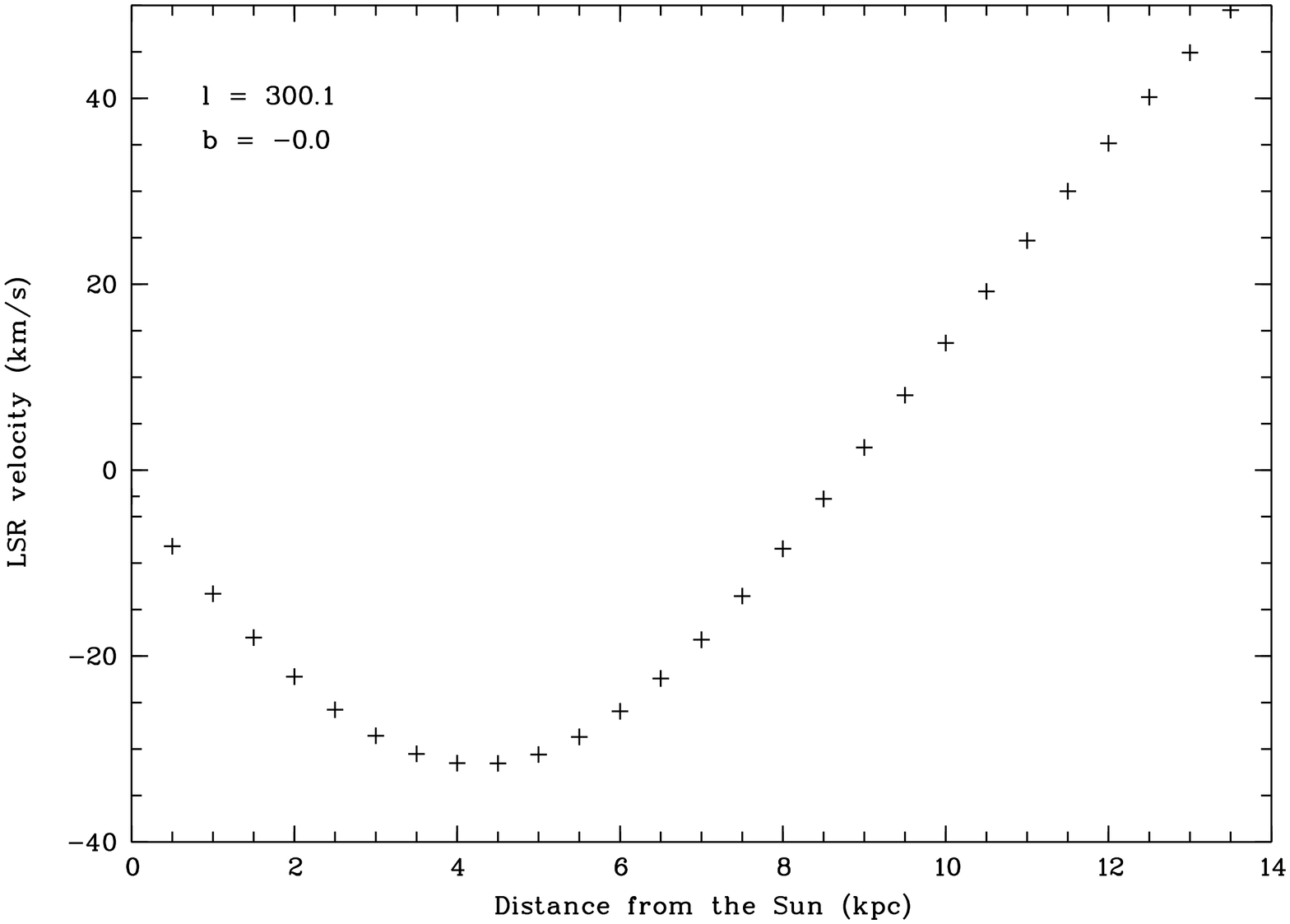}
\caption[]{{\it Top:} Interstellar $\ion{Na}{i}$ D doublet
          (5889.950~\AA\ full line, 5895.924~\AA\ dotted line) in a CAT/CES
          spectrum of \wray. The centroid velocities of the different
	  components (with respect to the local standard of rest) are
	  indicated; also the edge velocities of the central component
	  are given. {\it Bottom:} Radial velocity with respect to the
	  local standard of rest (LSR) due to galactic rotation as a
	  function of distance in the direction of \wray. A minimum of
	  $-31.5$~\kms\ is reached at a distance of about 4.5~kpc.}
\label{figsodium}
\end{center}
\end{figure*}

\section{Constraints on distance and mass of \wray}
\label{constraints}

\subsection{The interstellar $\ion{Na}{i}$ D lines}
\label{na_d_lines}

In principle, a lower limit to the distance of \wray\ can be derived
from a study of the (radial) velocity distribution of the interstellar
material in the direction towards \wray\ ($l^{II}=300 \fdg 1$;
$b^{II}=-0 \fdg 0$). For this purpose we use the interstellar
$\ion{Na}{i}$ D lines (5889.950~\AA, 5895.924~\AA) shown in
Fig.~\ref{figsodium}. The velocity scale is with respect to the local
standard of rest (LSR, 16.6~\kms\ towards right ascension 17:49:58.7
and declination +28:07:04, epoch 2000.0).

Three absorption components are detected in both transitions of the
$\ion{Na}{i}$ resonance doublet: one component centered at a velocity of
$+0.31$~\kms, a broad component with a width of $\sim 35$~\kms\ at
$-29.5$~\kms\ and a narrow component at $-56.4$~\kms. Assuming
circular galactic rotation and adopting the galactic rotation curve of
\citet{Brand93} with 8.5~kpc for the distance to the galactic center
and a galactic rotation velocity of 220~\kms\ at the orbit of the Sun,
the radial velocity along the line of sight towards \wray\ due to the
differential galactic rotation can be calculated as a function of
distance (Fig.~\ref{figsodium}). In this direction the radial velocity
with respect to the LSR reaches a minimum of $-31.5$~\kms\ at a
distance of about 4.5~kpc.

The line of sight in the direction of \wray\ first passes the Southern
Coalsack and intersects the Sagittarius-Carina spiral arm twice. The
longitude-velocity diagram of CO emission integrated over latitude
\citep{Cohen85} shows that the near intersection with the
Sagittarius-Carina arm yields CO emission in the velocity range
$\sim [-45:0]$~\kms; the far intersection with the Sagittarius-Carina
arm (outside the solar circle at a distance of at least 10~kpc)
results in CO emission at positive LSR velocities (see also
\citet{Crawford91} for $\ion{Na}{i}$ D observations). The absence of
$\ion{Na}{i}$ D absorption components at positive velocities indicates
that \wray\ must be located closer than the far intersection with the
Sagittarius-Carina arm. The velocity range covered by the broad
$\ion{Na}{i}$ D absorption component centered at $-29.5$~\kms\ is in very
good agreement with the CO emission produced by the near intersection
with the Sagittarius-Carina arm ($d \sim 1-2.5$~kpc). We note,
however, that the galactic rotation curve predicts velocities in the
range $[-25:-10]$~\kms\ at these distances. The more negative velocities
observed in $\ion{Na}{i}$ D absorption, especially the component at
$-56.4$~\kms, are difficult to explain with this rotation curve. This
is not very surprising as the radial velocity residuals between the
observations and model predictions of \citet{Brand93} in
this direction are on the order of 10~\kms. 

Several studies of the spiral-arm structure in this area of the sky
have led to the suggestion that at a distance of about 3-4~kpc
another spiral arm is present, the Norma-Centaurus arm
\citep{Bok71,Chukit73,Sembach93}. This arm would give rise to
absorption in the velocity range \mbox{$[-50:-30]$~\kms}. The Norma-Centaurus
arm is well defined by the $\ion{H}{ii}$ regions it contains
\citep{Courtes72}, but is less readily distinguished in $\ion{H}{i}$
surveys \citep{Kerr86} and CO emission in this direction
\citep{Cohen85,Dame87}. The lines of sight towards several stars in
this direction:

\noindent
\begin{center}
\begin{tabular}{lll}
HD103779 & $l^{II}=296 \fdg 9$, $b^{II}=-1 \fdg 0$ & $d \sim 4.1$~kpc \\
HD104683 & $l^{II}=297 \fdg 7$, $b^{II}=-2 \fdg 0$ & $d \sim 4.6$~kpc \\
HD113012 & $l^{II}=304 \fdg 2$, $b^{II}=+2 \fdg 7$ & $d \sim 4.1$~kpc \\
HD116852 & $l^{II}=304 \fdg 9$, $b^{II}=-16 \fdg 3$ & $d \sim 4.8$~kpc
\\
\end{tabular}
\end{center}

\noindent
yield remarkably similar $\ion{Na}{i}$ D spectra as observed for \wray\
\citep{Sembach93}. All these spectra include, besides the components
at 0 and $-25$~\kms, a narrow component around
$-50$~\kms. \citet{Sembach96} interpret this component as being due to
absorption in the Norma-Centaurus spiral arm.

\citet{Crawford91} give an alternative explanation for this narrow
component at $-50$~\kms. They derive an estimate of the distance to
the WN8 star We~21 which is located in the same region on the sky as
\wray\ ($l^{II}=302 \fdg 2$;$b^{II}=-1 \fdg 3$) using the method
described above. They do detect $\ion{Na}{i}$ D absorption at positive
velocities in the spectrum of We~21 and conclude that We~21 must be
located in the far part of the Sagittarius-Carina spiral
arm. \citet{Crawford91} show a figure of the interstellar
$\ion{Na}{i}$~D$_{2}$ line observed towards HD111904, a member of the
Cen~OB1 association in the Sagittarius-Carina arm at a distance of
2.5~kpc \citep{Humphreys78}. This line of sight ($l^{II}=303 \fdg
2$;$b^{II}=+2 \fdg 5$) has been discussed by \citet{Crawford92}. Just
like in the spectrum of \wray\ the $\ion{Na}{i}$~D (and
$\ion{Ca}{ii}$~K) lines towards HD111904 contain narrow and
blue-shifted high-velocity components (at $-30$ and $-45$ \kms). The
relative galactic rotation velocity at this distance is $-26$~\kms,
consistent with the LSR velocity of Cen~OB1
\citep[$-23.8$~\kms]{Humphreys72}, indicating that the gas forming
these narrow components has a substantial blue-shifted velocity with
respect to Cen~OB1. \citet{Crawford92} argues that the narrow
component at $-45$ \kms\ originates in an expanding shell surrounding
Cen~OB1, which is swept up by the cumulative action of the stellar
winds from the OB~stars in the association. The low value observed for
the $\ion{Na}{i}$/$\ion{Ca}{ii}$ line ratio of this component supports
this interpretation (Ca atoms are removed from grain surfaces by
intermediate-velocity shocks). The origin of the component at
$-30$~\kms\ might be a diffuse interstellar cloud with a peculiar
velocity of 5~\kms.
\begin{table}[!t]
\caption[]{Orbital parameters of \object{GX301$-$2} obtained by
          combining the results of \citet{Sato86} and BATSE
          \citep{Koh97}, corresponding to the values listed in
          \citet{Bildsten97}: the orbital period $P_{\rm orb}$,
          projected semi-major axis of the pulsar's orbit $a_{\rm X}
          \sin{i} $, the radial-velocity amplitude $K_{\rm X}$, the
          eccentricity $e$, the periastron angle $\omega$ and mass
          function $f(M_{\rm opt})$.}
\begin{center}
\begin{tabular}{ll}
\hline \hline
parameter & value \\
\hline
$P_{\rm orb}$ (days) & $41.498 \pm 0.002 $ \\
$a_{\rm X} \sin{i} $ (lt-s) & $368.3 \pm 3.7 $ \\
$K_{\rm X}$          & $218.3 \pm 3.3$ \\
$e$                  & $0.462 \pm 0.014$ \\
$\omega$ (deg)       & $310.4 \pm 1.4$ \\
$f(M_{\rm opt})$ (M$_{\sun}$) & $31.1 \pm 0.9$ \\  
Time of periastron passage (MJD) & $48802.79 \pm 0.12$ \\
\hline
\end{tabular}
\end{center}
\label{taborbit}
\end{table} 

The narrow $\ion{Na}{i}$~D absorption component in the spectrum of \wray\
centered at $-56.4$~\kms\ might as well be due to the expanding shell
around Cen~OB1. More speculative, (part of) the narrow component at
$0.3$~\kms\ could be formed by the receding part of the expanding
shell: the median velocity of the two narrow absorption components is
$-28.3$~\kms, not much different from the LSR velocity of
Cen~OB1. Alternatively, this low-velocity absorption could be due to 
the nearby Southern Coalsack \citep[$d \sim 170$~pc,][]{Franco89}.
\begin{figure*}[!t]
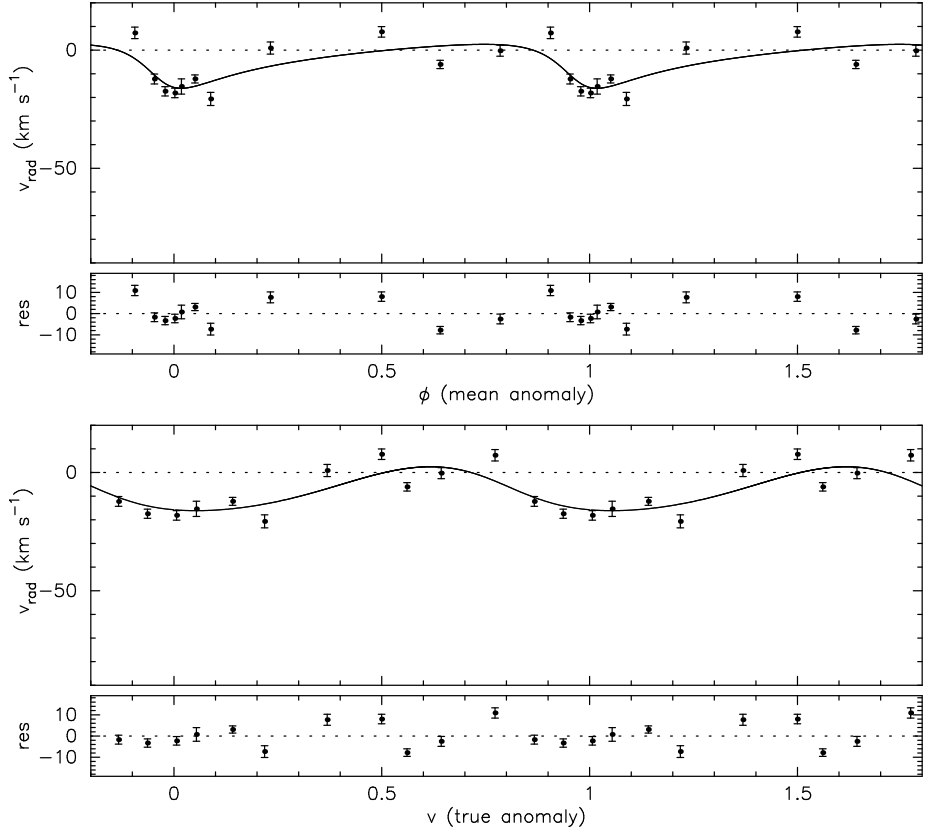

\begin{center}
\includegraphics[width=12.2cm]{5393f06a.ps}
\\[0.2cm]
\includegraphics[width=12.2cm]{5393f06b.ps}
\caption[]{Radial velocity curve of the $\ion{Si}{iii}$ line at
          4813.33~\AA\ of \wray, as a function of mean anomaly ({\it
	    top}) and true anomaly ({\it bottom}). The fit shows a
	  curve with $K_{\rm opt} = 9.3 \pm 2.7$~\kms, $v_{\gamma} =
	  -4.1 \pm 2.4$~\kms\ and $\triangle \phi = 0.04 \pm 0.04$;
	  the reduced $\chi^{2}$ of this fit is 9.8. Apparently, the
	  1$\sigma$-errors on the radial-velocity measurements are
	  much smaller than the deviation from the Keplerian orbit.}
\label{figradvel}
\end{center}
\end{figure*}

On the basis of the $\ion{Na}{i}$~D interstellar lines we come to the
conclusion that the distance to \wray\ is at least 2.5~kpc,
i.e. beyond the near intersection with the Sagittarius-Carina arm. The
large colour excess, the strong $\ion{Na}{i}$~D absorption (e.g. the
members of the OB association Cen~OB1 in the Carina arm at a distance
of 2.5~kpc have a much smaller colour excess \citep{Humphreys78} and
also their $\ion{Na}{i}$~D absorption is much weaker) and the
$\ion{Na}{i}$~D component at $-56.4$~\kms\ suggest that \wray\ is
located in (or behind) the Norma-Centaurus arm, which has a distance
of $3-4$~kpc. The distance estimates of 5.3~kpc proposed in
\citet{Kaper95} and 7.7~kpc in \citet{VanGenderen96} most likely
overestimate the true distance to \wray.
\begin{table}[!t]
\caption[]{Radial-velocity curves obtained for individual lines in the
          VLT/UVES spectra of \wray. The high value of $\chi_{r}^{2}$
          indicates that the errors on the individual radial-velocity
          measurements are much smaller than the measured deviations
          from the best-fit curve.}
\begin{center}
\begin{tabular}{lllll}
\hline \hline
Line $\lambda$ & $K_{\rm opt}$ & $v_{\gamma}$ & $\triangle \phi$ & $\chi_{r}^{2}$ \\ 
(\AA )         & (\kms )       & (\kms )      &                  &                \\ 
\hline
4366.89 & $10.9 \pm 2.6$ & \0$-7.3 \pm 2.4$  & $0.01 \pm 0.02$ & \035.6  \\
4661.63 & $10.2 \pm 2.7$ & \0$-7.7 \pm 2.4$  & $0.02 \pm 0.03$ & 105.1 \\
4813.33 & \0$9.3 \pm 2.7$  & \0$-4.1 \pm 2.4$  & $0.03 \pm 0.03$ & \0\09.8   \\
4819.72 & \0$9.6 \pm 3.3$  & \0$-3.1 \pm 3.0$  & $0.04 \pm 0.04$ & \026.5  \\
4828.96 & $10.6 \pm 2.9$ & \0$-1.8 \pm 2.7$  & $0.03 \pm 0.03$ & \027.1  \\
5686.21 & \0$9.5 \pm 3.5$  & $-19.7 \pm 2.8$ & $0.00 \pm 0.03$ & 135.5 \\
5696.60 & $10.2 \pm 3.9$ & $-13.3 \pm 3.1$ & $0.01 \pm 0.04$ & 291.8 \\
5722.73 & $10.2 \pm 3.8$ & $-10.6 \pm 3.2$ & $0.02 \pm 0.03$ & 119.7 \\
5739.73 & $11.0 \pm 3.1$ & \0$-9.3 \pm 2.6$  & $0.00 \pm 0.03$ & 377.3 \\
\hline
\end{tabular}
\end{center}
\label{tabradvel}
\end{table}

\subsection{Radial-velocity curve of \wray}
\label{radvel}

A constraint on the mass of the two binary components is obtained from
the radial-velocity curve of \wray. \citet{Hutchings82} measured
radial velocities for \wray\ and arrived at a radial-velocity
amplitude $K_{\rm opt} = 18 \pm 4$~\kms\ and a $\gamma$-velocity of $-31 \pm
2$~\kms\ for a (then not confirmed) orbital period of 41.37d and using
photospheric lines\footnote{\citet{Hutchings82} report
a systematic dependence of the derived radial velocity on hour angle
for which they correct; however, in the update following their paper
to comment on the newly derived orbital period they do not find an
hour-angle effect.}. 

We measured the radial velocity of \wray\ from the photospheric lines
included in the CES and UVES spectra by fitting gaussians to the
spectral line profiles. The radial velocity listed in
Tables~\ref{tablogces} and \ref{tabloguves} is the mean radial
velocity per spectrum; the error corresponds to the standard
deviation. In the CES spectra four $\ion{N}{ii}$ lines (at 5666.63,
5676.02, 5679.56 and 5686.21~\AA) were fit with a gaussian to measure
the radial velocity (Table~\ref{tablogces}). As only a very narrow
orbital phase interval was covered (periastron passage), it is not
possible to derive $v_{\gamma}$, $\Delta \phi$ and $K_{\rm opt}$ from
this dataset. The observed trend in $v_{\rm rad}$ is, however,
consistent with the radial velocity curve obtained from the UVES
spectra (Table~\ref{tabloguves}). In the UVES spectra nine lines were
measured (Table~\ref{tabradvel}). A radial-velocity curve is obtained
for each individual line; we selected only weak photospheric lines
that are not disturbed by variations in the stellar wind (see
Sect.~\ref{interaction}). As an example Fig.~\ref{figradvel} shows
the results obtained for the $\ion{Si}{iii}$ line at 4813.33~\AA, for
which we obtained the best fit. Subsequently, the radial-velocity
curves were fit by a model based on the orbital parameters of the
X-ray pulsar (Table~\ref{taborbit}), with the radial-velocity
amplitude $K_{\rm opt}$, the $\gamma$-velocity and a phase shift
$\triangle \phi$ as free parameters (Table~\ref{tabradvel}). The high
values for $\chi_{r}^{2}$ indicate that the errors on the individual
radial-velocity measurements are much smaller than the measured
deviations from the best-fit Keplerian orbit.

These radial-velocity excursions are systematic (i.e. similar for each
line in a given spectrum) and are also observed in other eccentric
HMXBs, e.g. \object{Vela~X$-$1} \citep{VanKerkwijk95,Barziv01}. A possible
physical origin of these radial-velocity excursions are tidally forced
non-radial oscillations in the OB companion. \citet{Quaintrell03} find
evidence for a 2.18~d periodicity in the residuals of the Keplerian
fit to the 8.96~d orbit, and use this to obtain a more accurate
determination of $K_{\rm opt}$. A similar effect may be underlying the
radial-velocity curve of \object{Wray~977}, but we have too few points
to measure it, if present. \citet{VanGenderen96} report a 11.90~d
quasi-period in photometry of \object{Wray~977}, as for
\object{Vela~X$-$1} about 1/4th of the orbital period. Having said
this, inspection of Table~\ref{tabradvel} shows that the measured
radial-velocity amplitude is about the same for each line. We
conservatively adopt $K_{\rm opt} = 10 \pm 3$~\kms. For a more
detailed description of the used method we refer to
\citet{VanderMeerPhD,VanderMeer06}.


With the measured radial-velocity amplitude of \wray\ and the orbit of
the X-ray pulsar (Table~\ref{taborbit}) the mass ratio $q =
M_{X}/M_{\rm opt}$ of the system is determined:
\begin{equation}
q = \frac{K_{\rm opt}}{K_{X}} = \frac{10 \pm 3}{218.3} = 0.046 \pm 0.014
\end{equation}
The mass function based on the X-ray pulsar's orbit is:
\begin{equation}
f(M_{\rm opt}) = \frac{M_{\rm opt} \sin^{3}{i}}{(1 + q)^{2}} = 31.1 \,
{\rm M}_{\sun}
\end{equation}
Note that \citet{Koh97} quote a mass function of 31.8~M$_{\sun}$, but
this value is based on the ephemeris of \citet{Sato86}. Using the
values listed in Table~\ref{taborbit} one obtains 31.1~M$_{\sun}$
\citep[see also][]{Bildsten97}. For the conservative range of $7 < K_{\rm
opt} < 13$~\kms\ we get $0.032 < q < 0.060$. The absence of an X-ray
eclipse (but see Sect.~\ref{interaction}) provides an upper limit to
the inclination $i$ of the system's orbit:
\begin{equation}
\tan{i} < 1.77 \times 10^{2} \left( \frac{1 + q}{R_{\rm opt}} \right)
\end{equation}
with $R_{\rm opt}$ the radius of \wray\ (in $R_{\odot}$). This
parameter is difficult to determine; in Sect.~\ref{modelling} we use
model-atmosphere fits to the optical spectrum to derive a measure of
the stellar radius ($R_{*} = 62$~R$_{\sun}$). Clearly, $R_{\rm opt}$
should not become bigger than the distance to the X-ray pulsar at
periastron passage. Also, we will consider whether \wray\ fits into
its ``Roche-lobe'', even though this constraint is not valid in this
eccentric system. The Roche-lobe radius $R_{L}$ is smaller than the
so-called tidal radius $R_{T}$; for a definition, see, e.g.,
\citet{Joss84}. We will show that the derived radius of \wray\ is of
the order of (and probably even larger than) $R_{L}$ and $R_{T}$ at
periastron passage. In the following analysis we leave $R_{\rm opt}$
as a free parameter.
\begin{figure*}[!t]
\begin{center}
\includegraphics[angle=-90,width=11.2cm]{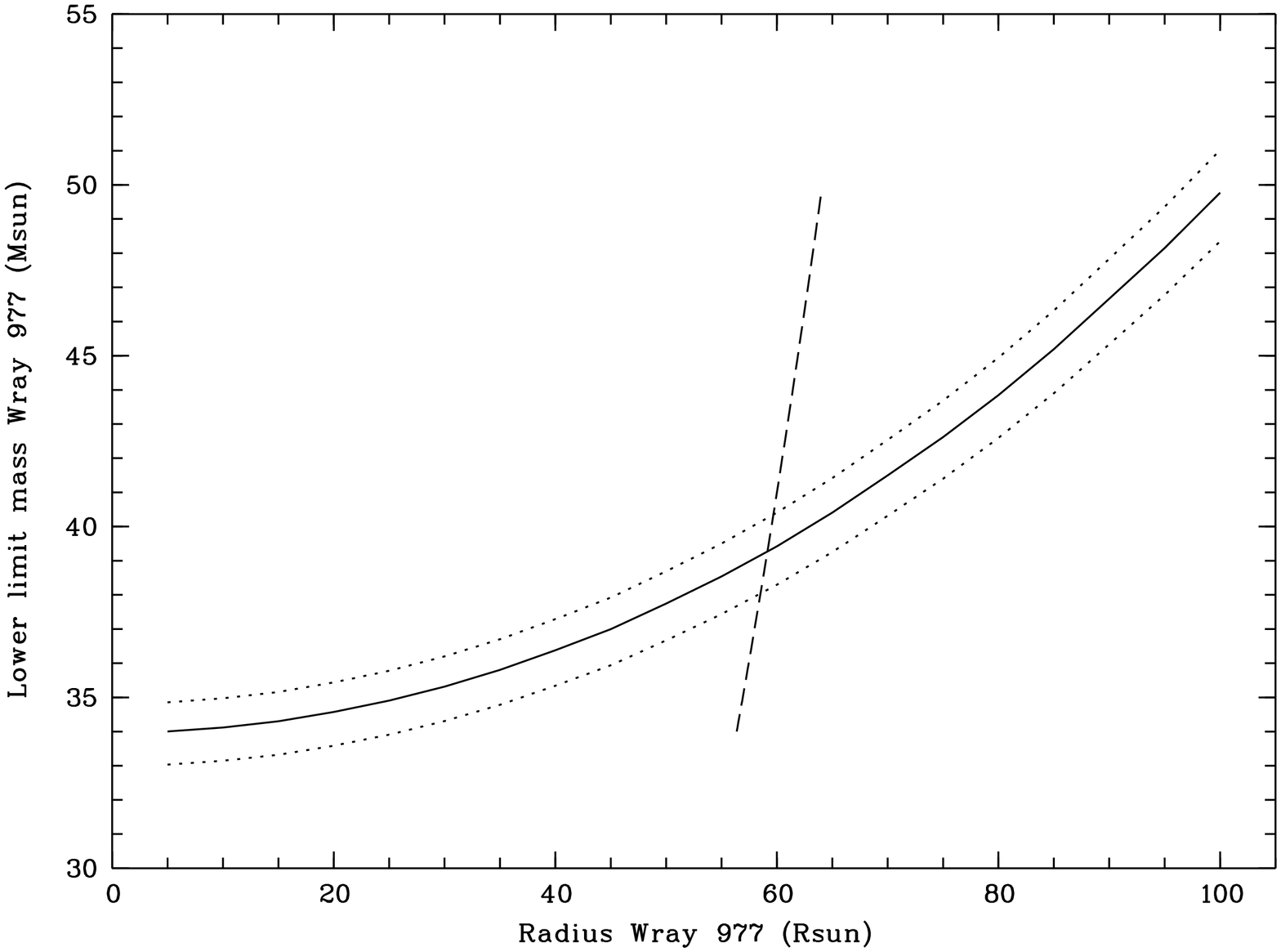}
\\[-0.2cm]
\includegraphics[angle=-90,width=11.2cm]{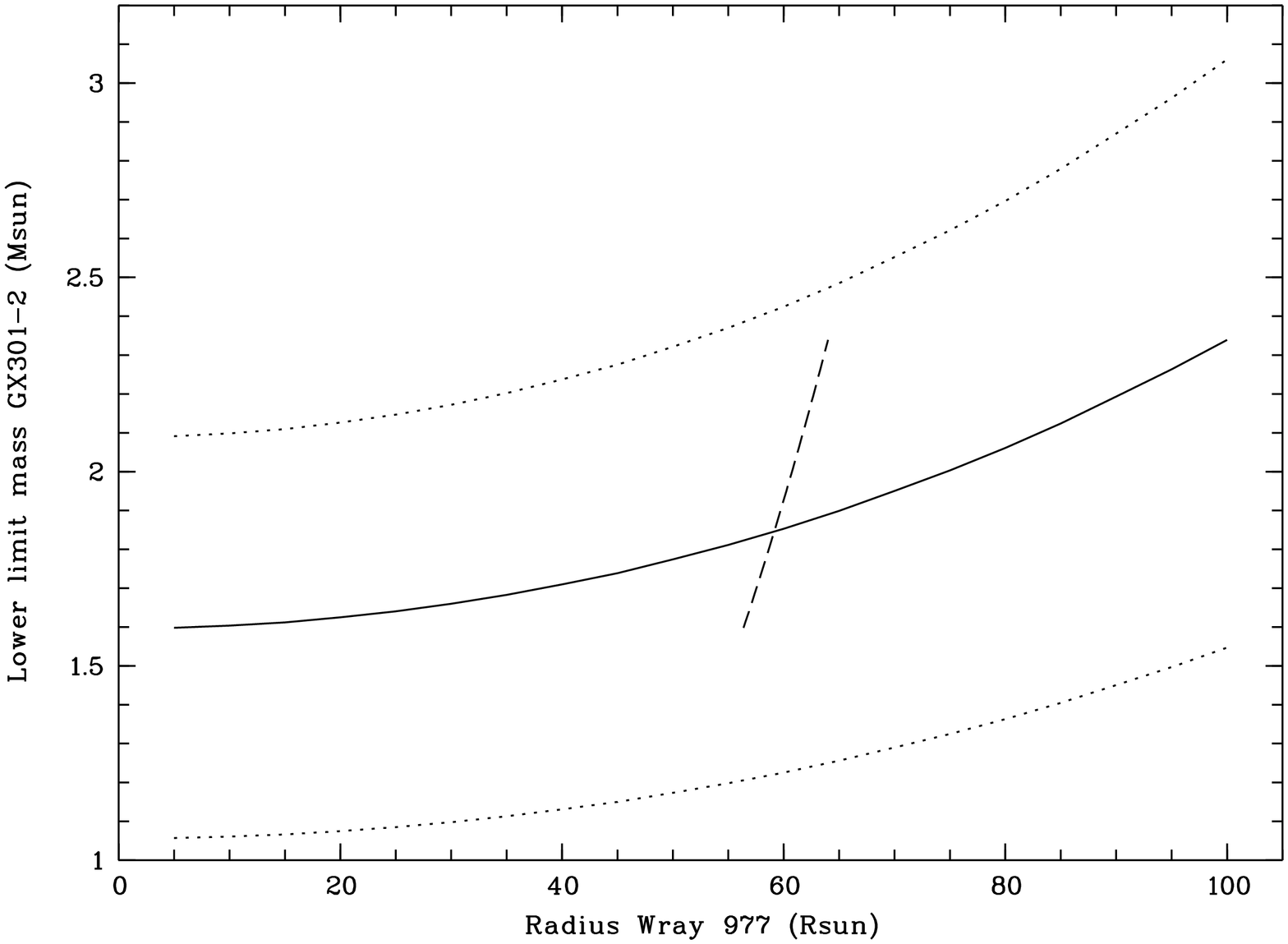}
\caption[]{Lower limit on the mass of \wray\ ({\it top}) and on the
          mass of \object{GX301$-$2} ({\it bottom}) as a function of
          the radius of \wray. The drawn line is obtained for $q =
          0.046$; the dotted lines are for $q = 0.032$ and $q = 0.060$
          for the lower and upper line, respectively. The vertical (dashed)
          line represents the size of the Roche-lobe radius $R_{L}$ at
          periastron passage.}
\label{figradwray}
\end{center}
\end{figure*} 

From the above we can set lower limits to the masses of \wray\ and
\object{GX301$-$2} and investigate how these depend on $R_{\rm opt}$
(Fig.~\ref{figradwray}). The figures show three
lines, each for a different value of the radial velocity amplitude:
the best value is $K_{\rm opt} = 10$~\kms\ (drawn line, $q=0.046$);
the dotted lines correspond to $q = 0.032$ and $q = 0.060$. From
Fig.~\ref{figradwray} we conclude that the mass of \wray\ is higher than
$34 \pm 1$~M$_{\sun}$. Given a radius of 62~R$_{\sun}$, the lower
limit on the mass becomes $39 \pm 1$~M$_{\sun}$. The corresponding
upper limit on the inclination $i$ is 72 degrees. Note that the mass of
\wray\ is inversely proportional to $\sin^{3}{i}$; e.g. for $i =
30^{\circ}$ M$_{\rm opt}$ would become 272~M$_{\sun}$.

Fig.~\ref{figradwray} indicates that the lower limit on the mass of the
neutron star is 1.6~M$_{\sun}$ if $K_{\rm opt}$ is 10~\kms. The
uncertainty on the neutron-star mass is large, as can be seen from the
figure. For $R_{\rm opt} = 62$~R$_{\sun}$ the lower limit on the mass
of \object{GX301$-$2} is $1.85 \pm 0.6$~M$_{\sun}$.
\begin{table}[!t]
\caption[]{Lower limit on the system inclination and upper limit to
          the mass of \wray\ set by the upper limit to the mass of the
          neutron star \object{GX301$-$2}: 3.2~M$_{\sun}$ and
          2.5~M$_{\sun}$ (in brackets) and a radius of 62~R$_{\sun}$
          for \wray.}
\begin{center}
\begin{tabular}{lll}
\hline \hline
$q$ &  Lower limit $i$                & Upper limit M$_{\rm opt}$ \\
    &  M$_{X} < 3.2 (2.5)$~M$_{\sun}$ & (M$_{\sun}$)              \\
\hline
 0.032 & 43.7 (48.6) & 100 (78) \\
 0.046 & 52.5 (59.4) & 68 (53)  \\
 0.060 & 60.2 (70.4) & 53 (42)  \\
\hline
\end{tabular}
\end{center}
\label{tabmwray}
\end{table}

The neutron star cannot be more massive than 3.2~M$_{\sun}$, a limit
set by general relativity \citep{Nauenberg73,Rhoades74,Kalogera96}. It
is likely that the maximum neutron star mass is determined by the
stiffness of the equation of state of matter at supra-nuclear
density. Then the maximum mass would be about 2.5~M$_{\sun}$ (for a
detailed discussion on the maximum neutron-star mass see
\citealt{Srinivasan02}). The currently most massive X-ray pulsar is
Vela~X-1 with a mass of $1.86 \pm 0.16$~M$_{\sun}$
\citep{Barziv01,Quaintrell03}. With an upper limit to the neutron-star
mass we can set a {\it lower} limit to the system inclination. This
implies an {\it upper} limit to the mass of \wray\
(Table~\ref{tabmwray}).

Thus, for a radius of 62~R$_{\sun}$ and a mass ratio $q = 0.046$ the
mass of \wray\ is in the range $39 < {\rm M}_{\rm opt} < 68
(53)$~M$_{\sun}$ (in brackets the mass for a maximum neutron star
mass of 2.5~M$_{\sun}$).  Also, with high probability the inclination
of the system $i = 60 \pm 10$~deg. This conclusion is consistent with
the observed X-ray lightcurve (Sect.~\ref{interaction}).

\section{Spectrum modelling}
\label{modelling}

Several sophisticated, non-LTE atmospheric codes have been developed to
model the energy distribution and detailed spectral line profiles of
hot, luminous stars
\citep[e.g.,][]{Pauldrach01,Puls05,Dekoter93,Mokiem05,Aufdenberg02}.
We have utilised the iterative, non-LTE line blanketing method of
\citet{Hillier98} to perform our quantitative spectral analysis. The
code solves the radiative transfer equation in the co-moving frame for
the expanding atmospheres of early-type stars in spherical geometry,
subject to the constraints of statistical and radiative equilibrium.
Steady state is assumed and the density structure is set by the
stellar-wind mass-loss rate and velocity field via the continuity
equation.  The velocity field consists of a pseudo-hydrostatic
photosphere in the inner region characterised by the gravity
($\log{g}$) of the star with a smooth transition to the supersonic
wind region. The latter is parameterised in terms of a wind velocity
law $v(r)$ (``$\beta$ law'') and allows for the possibility to define
two different $\beta$ values to account for a second accelerating
zone.  Clumping of the stellar wind is included via a clumping law
characterised by a volume filling factor $f(r)$, so that the
``smooth'' mass-loss rate, $\dot{M}_{S}$, is related to the
``clumped'' mass-loss rate, $\dot{M}_{C}$, through $\dot{M}_{S} =
\dot{M}_{C}/ \sqrt{f}$.  The model is then prescribed by the stellar
radius $R_{*}$, the stellar surface gravity $\log{g}$, the stellar
luminosity $L_{\rm *}$, the mass-loss rate $\dot{M}$, the velocity
field $v(r)$, the volume filling factor $f$ and the abundances of the
considered chemical elements. The reader is referred to
\citet{Hillier98,Hillier99} for a detailed discussion of the code.
\begin{table}[!t]
\caption[]{Basic properties and derived model parameters of \wray.}
\begin{center}
\begin{tabular}{p{4cm}r}
\hline \hline
\multicolumn{2}{c}{Magnitudes and colour} \\
\hline
$V$		&	10.83		\\
$B-V$		&	$1.76$		\\
$E(B-V)$	&	$1.96$		\\
$M_V$		&	$-7.65$		\\
$d$ (kpc)	&	$3.04$		\\ 
\hline
\multicolumn{2}{c}{Model parameters} \\
\hline
$\log{(L_{*}/L{_{\sun}})}$              & 5.67 \\ 
$\log{g}$                               & 2.38 \\
$R_\star/R_{\sun}$                      & 62 \\
$R_{2/3}/R_{\sun}$                      & 70 \\
$T_\star$ ($10^4$K)                     & 19.1 \\
$T_{\rm eff}$ ($10^4$K)                 & 18.1 \\
$M_{*}/M_{\sun}$                        & 43 \\
$v(R_{2/3})$ (\kms)                     & 4.40 \\
$v \sin{i}$ (\kms)                      & 50   \\
$\log(\dot{M})$ (M$_{\sun}$~yr$^{-1}$)  & $-5.00$ \\
$v_\infty$ (\kms )                      & 305 \\
$\beta$                                 & 1.75 \\
$f$                                     & 1.0 \\
$\eta=\dot{M}v_\infty/(L_\star/c)$      & 0.32 \\
\hline
\end{tabular}
\end{center}
\label{modelpar}
\end{table}

\subsection{Stellar parameters}
\label{stellarparameters}

\begin{figure*}[!t]
\begin{center}
\includegraphics[width=15cm]{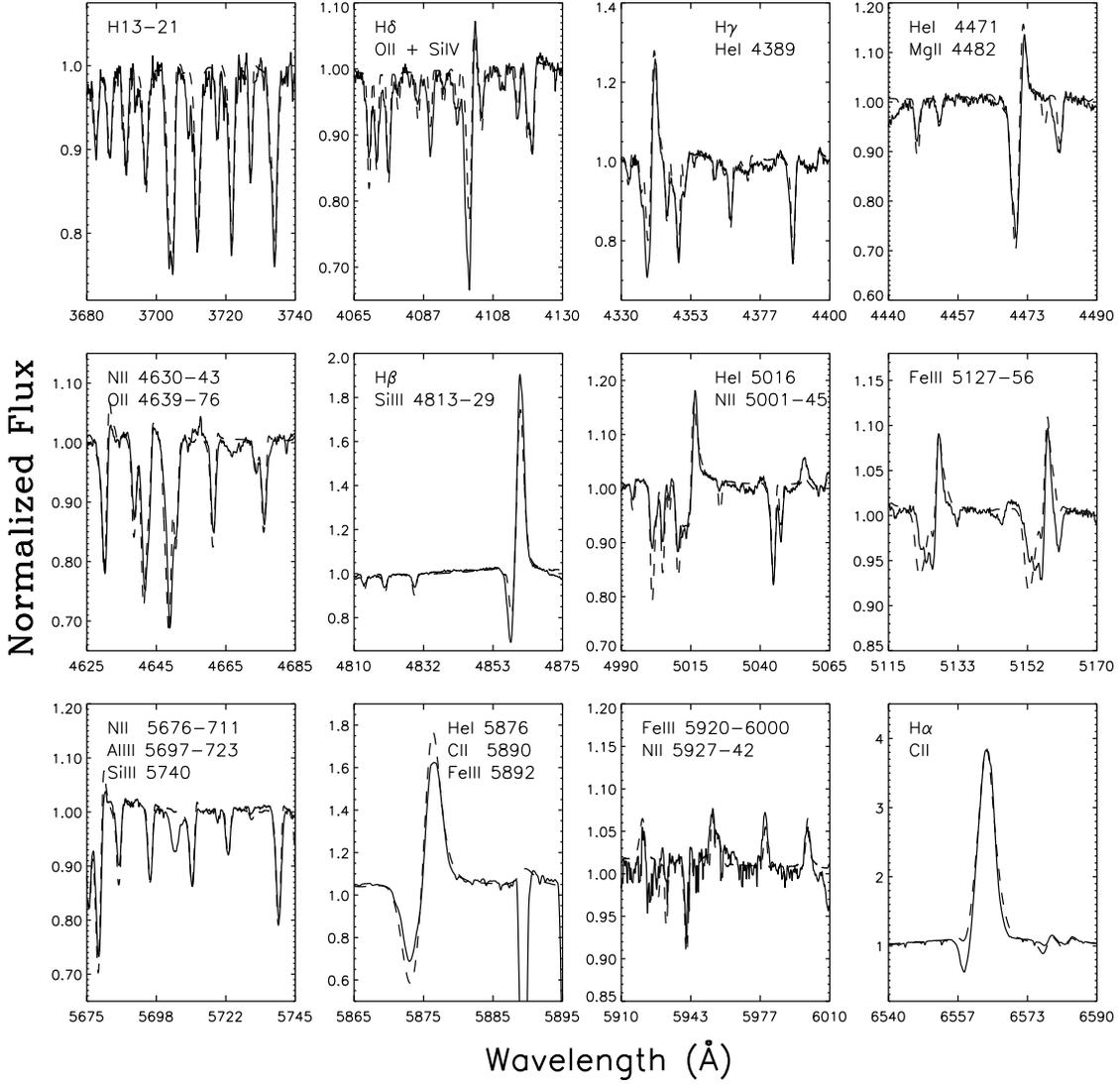}
\caption[]{Different spectral diagnostics in the spectrum of \wray\
          (at orbital phase $\phi=0.78$) compared to our best fit
          model (dashed line).}
\label{model}
\end{center}
\end{figure*}

Given the uncertainty in the distance to Wray 977 discussed in
Sect.~\ref{constraints}, we adopted a fixed value for the radius of
the star. We set $R_{2/3}=70$~$R_{\sun}$, which corresponds to
a distance of 3~kpc (see the discussion below). The main stellar
parameters and chemical abundances of our best-fitting model are
summarised in Tables~\ref{modelpar} and
~\ref{abundances}. Figure~\ref{model} illustrates the excellent
agreement of our model with some of the most important diagnostic
lines.

The parameters $L_\star$ and $T_\star$ in Table~\ref{modelpar} refer to the
stellar radius $R_\star$ at which the Rosseland optical depth reaches a
value of $\sim 30$, while the stellar effective temperature $T_{\rm
eff}$ is obtained at an optical depth of 2/3 ($R=R_{2/3}$).  From
Table~\ref{modelpar} we see that $v(R_{2/3})>4$~\kms, so basically no
information can be obtained from the hydrostatic region
(i.e. $\log{g}$) as the stellar wind has already set in at this
depth. Even the deeper forming wings of the higher members of the
Balmer series are dominated by the run of the density structure within
the transition region from the sub- to the supersonic atmosphere and
react only minorly to changes in $\log{g}$ as high as 0.25~dex.  The
really extended nature of the object given by the $R_{2/3}/R_\star \sim
1.13$ ratio clearly confirms its previous classification as a
hypergiant \citep{Kaper95}.

The effective temperature of the object is very well constrained by
the $\ion{Si}{iv}$/$\ion{Si}{iii}$ and $\ion{Si}{iii}$/$\ion{Si}{ii}$
ionisation balance. The presence of diagnostic lines of all three
ionisation stages of silicon allows to reach an accuracy of better
than 500~K. Some other transitions of species such as $\ion{N}{ii}$,
$\ion{O}{ii}$, $\ion{S}{iii}$ and $\ion{Fe}{iii}$, which react
strongly to changes in temperature (luminosity) in this parameter
domain, also show excellent agreement with $T_{\rm eff} = 18\,100$~K
determined from the silicon ionisation.

The stellar wind parameters also confirm the hypergiant nature of
\wray, with typical wind density and performance number $\eta$ well
below (factor 5 and 1.6, respectively) that of luminous blue variables
(LBVs) such as P~Cygni (\citet{Najarro97}, Najarro \& Hillier 2006, in
prep.). Finally, we note that the observed spectra of \wray\ do not
show clear evidence of clumping, as the electron scattering wings of
the Balmer lines are fairly well reproduced by a smooth wind
model. Further, the resulting spectra of our exploratory models
including clumping showed severe discrepancies with the high Balmer
series members and other lines formed in the inner atmosphere.
\begin{figure*}[!t]
\begin{center}
\includegraphics[width=12.2cm]{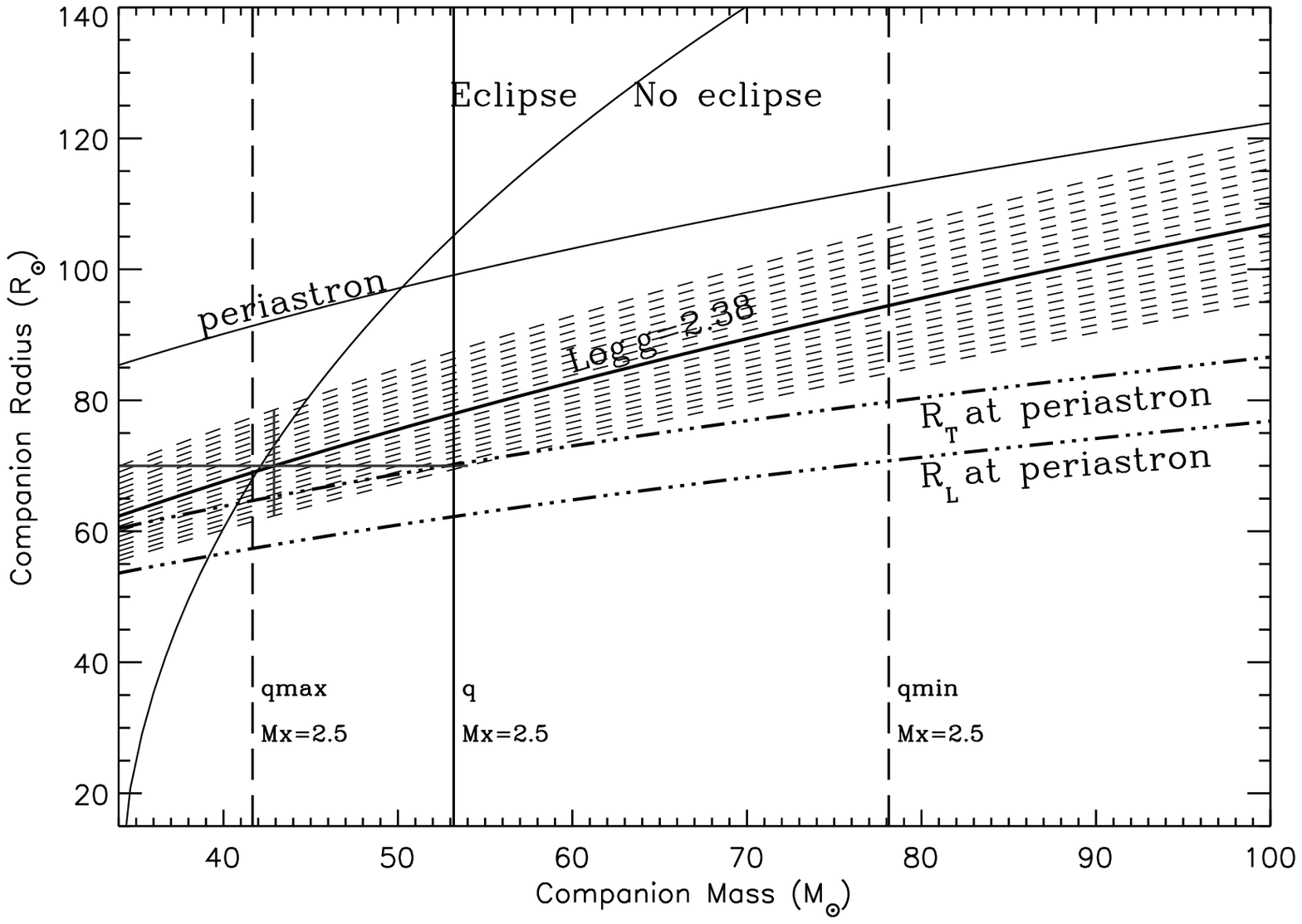}
\caption{Allowed domain for the mass of \wray\ as a function of its
        radius. The constraints are provided by the absence of an
        X-ray eclipse and the distance between \wray\ and the X-ray
        source at periastron passage. The tidal-lobe radius $R_T$ and
        Roche-lobe radius $R_L$ provide limits above which tidal and
        Roche-lobe interactions would occur, though the latter should
        be regarded with caution as the system is highly
        eccentric. Vertical lines provide limits on the mass based on
        the allowed range of the mass ratio $q$ and the adopted
        maximum neutron star mass (i.e. 2.5~M$_{\sun}$). For a radius
        of 70~$R_{\sun}$ the tidal interaction limit requires that
        M~$>53$~M$_{\sun}$ ($\log{g} > 2.47$). Tidal interaction would
        start about 1.5 days before periastron passage (i.e. during
        the X-ray flare event) and has a duration of 3 days if M~$\sim
        43$~M$_{\sun}$ ($\log{g}=2.38$).}
\label{eclipse}
\end{center}
\end{figure*}

Subsequently, we may use the derived stellar parameters together with
the information on the distance, radial-velocity curve and orbital
parameters of the system (Sect.~\ref{constraints}) to place some
harder limits on the radius and distance of the
star. Figure~\ref{eclipse} displays the allowed domain for the mass of
\wray\ as a function of stellar radius (similar to Fig.~10 in
\citealt{Koh97}). The constraints are provided by the absence of X-ray
eclipse and the distance between the star and the X-ray pulsar at
periastron passage (i.e., the stellar radius should not be larger than
the periastron distance). The tidal-lobe radius $R_T$ and Roche-lobe
radius $R_L$ provide limits above which tidal and Roche-lobe
interaction would occur, though the latter should be regarded with
caution as the system is highly eccentric. Vertical lines provide
limits on the mass of \wray\ when taking into account the allowed
range of the mass ratio $q$ (Sect.~\ref{radvel}) and the maximum
neutron-star mass (here we adopt a maximum mass of
2.5~M$_{\sun}$). For our best model ($R_{2/3} = 70 \, R_{\sun}$) the
tidal interaction limit requires that M$_{\rm opt} > 53$~M$_{\sun}$
($\log{g} > 2.47$). However, tidal interaction would start about
1.5~days before periastron passage (which roughly corresponds to the
timing of the X-ray flare, see Sect.~\ref{interaction}) with a duration
of 3 days if M$_{\rm opt} \sim 43$~M$_{\sun}$ ($\log{g}=2.38$).

On the other hand, if we require a lower limit on the distance of
2.5~kpc (see Sect.~\ref{na_d_lines}), then $R_{2/3} > 57.6 \,
R_{\sun}$ and M$_{\rm opt} > 39.6$~M$_{\sun}$ ($\log{g} > 2.51$).
With $q=0.046$ this would correspond to a neutron-star mass $M_{X} >
1.8$~M$_{\sun}$ and \wray\ would remain within its tidal lobe
throughout the orbit, i.e. no tidal interaction would occur. The main
drawback is the relatively high value of $\log{g}$ which is almost
0.2~dex above the expected value for a hypergiant.  If we assume that
the distance to the system is at least 3.5~kpc, then the resulting
stellar radius would be larger than 81~$R_{\sun}$ and M$_{\rm opt} >
45$~M$_{\sun}$ ($\log{g} > 2.27$). For $q=0.046$ we would then have
M$_{X} > 2.1$~M$_{\sun}$. While these values are appropriate for a
hypergiant, it would exceed its tidal lobe during 5.6 days (7~\% of
the orbit) starting 2.8 days before periastron passage.  Requiring no
tidal interaction would imply that M$_{\rm opt} > 84$~M$_{\sun}$
($\log{g} > 2.54$), which would be at odds with both the spectral type
and the maximum $q$ value (0.065) corresponding to a neutron star mass
of 2.5~$M_{\sun}$.

The values of the stellar parameters of \wray\ quoted in
\citet{Wellstein99}, $17,000 \leq T_{\rm eff} \leq 20\,000$~K and $60
\leq R \leq 70$~R$_{\sun}$, are compatible with our currently
preferred values: $T_{\rm eff} = 18,100 \pm 500$~K, $R_{\rm 2/3} =
70$~R$_{\sun}$. The corresponding luminosity $\log{(L_{*}/L{_{\sun}})}
= 5.67$ of \wray, adopting a distance of 3.0~kpc, is in good agreement
with the evolutionary track predicted by \citet{Wellstein99}. If one
would require that the radius of \wray\ is smaller than the tidal
radius {\it throughout} the orbit, $M_{\rm opt}$ should be at least
53~M$_{\sun}$, implying a high neutron-star mass: $M_{X} = 2.4 \pm
0.7$~M$_{\sun}$ for $q=0.046 \pm 0.014$. If one would allow \wray\ to
exceed its tidal lobe during periastron passage, $M_{\rm opt}$ may
become less. Therefore, we conclude that the mass interval derived for
\wray\ in Sect.~\ref{radvel}: $39 < M_{\rm opt} < 53$~M$_{\sun}$ is in
accordance with our model atmosphere results (and a not too high
neutron-star mass), noting that tidal interaction is expected during
periastron passage (see Sect.~\ref{interaction}).
\begin{table}[!t]
\caption[]{Surface chemical abundances of \wray\ resulting from the
           best-fitting model atmosphere.}
\begin{center}
\begin{tabular}{llll}
\hline \hline
Species	& X/He		& Mass		& $X/X_{\sun}$ \\
	& (number)	& fraction	& \\ 
\hline
H	& $ 3.5	$               & $ 4.582 \times 10^{-1}$ & $6.5 \times 10^{-1}$ \\
He	& $ 1.0	$               & $ 5.236 \times 10^{-1}$ & $1.9$ \\
C	& $ 3.9 \times 10^{-3}$	& $ 6.107 \times 10^{-3}$ & $2.0$ \\
N	& $ 2.9 \times 10^{-3}$	& $ 5.250 \times 10^{-3}$ & $4.8$   \\
O	& $ 1.8 \times 10^{-3}$	& $3.815 \times 10^{-3}$  & $4.0 \times 10^{-1}$ \\
Mg	& $ 1.5 \times 10^{-4}$	& $4.867 \times 10^{-4}$  & $7.5 \times 10^{-1}$ \\
Al	& $ 1.6 \times 10^{-5}$ & $5.615 \times 10^{-5}$  & $1.0$ \\
Si	& $ 1.9 \times 10^{-4}$ & $6.988 \times 10^{-4}$  & $1.0$ \\
S	& $ 8.7 \times 10^{-5}$ & $3.673 \times 10^{-4}$  & $1.0$ \\
Fe	& $ 1.9 \times 10^{-4}$	& $1.356 \times 10^{-3}$  & $1.0$ \\ 
\hline
\end{tabular}
\end{center}
\label{abundances}
\end{table}

\subsection{Photospheric abundances}

Table~\ref{abundances} lists the chemical abundances obtained
from our best-fitting model. The derived H/He ratio (3.5 by number)
is consistent with the enrichment displayed in B hypergiants and is
very close to the characteristic value of LBVs (\citet{Najarro97},
Najarro \& Hillier 2006, in prep.).  From our analysis, the observed
ratio of H to $\ion{He}{i}$ lines confines the He abundance in the range
$2.5 \leq {\rm H/He} \leq 5$ (by number).

Apart from CNO elements, the derived surface abundances are fully
consistent with solar metallicity. Our models show a nitrogen
enrichment of roughly 5 times over solar and an oxygen depletion of
2.5 times below solar.  These values are fairly robust, as the number
of diagnostic lines for both elements is relatively large and display
excellent internal agreement.  These two values should be regarded
with an accuracy better than a factor of two.  For carbon we obtain an
enhancement twice over solar. Unfortunately, only two $\ion{C}{ii}$
lines are available for abundance analysis (4267~\AA\ and the doublet
6578--82~\AA ) and both react very strongly to the density structure
within the sub- to supersonic transition region.  Thus the uncertainty
on the carbon abundance can be regarded as high as at least a factor
2. The derived nitrogen enhancement and oxygen depletion agree very
well with the surface abundances predicted by Model nr.~8 of
\citet{Wellstein99}. Their conservative binary evolution model for
\wray\ predicts a nitrogen enrichment by a factor of 5.5 and an oxygen
depletion by a factor 0.75. Also carbon should be depleted by a factor
0.33, which is not indicated by our spectral analysis. In a fully
non-conservative binary evolution history (i.e., mass has not been
transferred from the primary to the secondary, but has been lost from
the system) the chemical surface abundances of \wray\ should not have
been affected.

\subsection{Projected rotational velocity}

The projected rotational velocity of \wray\ derived from the spectral
line fits is $v \sin{i} = 50 \pm 10$~\kms. This corresponds to a
maximum rotation period of \wray\ of 63~days, given a radius of
62~$R_{\sun}$. Simulations of tidal mass transfer in a system like
\wray\ by \citet{Layton98} show that if the primary star is rotating
near corotation with the orbiting compact object at periastron,
tidally stripped gas can accrete and cause X-ray flares. The measured
value of $v \sin{i}$ indicates that this condition can only be met if
the inclination of the system $i \la 10^{\circ}$. We can exclude
corotation at periastron passage as at such a low value of the orbital
inclination the mass of the neutron star exceeds the causality limit
by far (Sect.~\ref{radvel}).
\begin{figure*}[!t]
\begin{center}
\includegraphics[width=12cm]{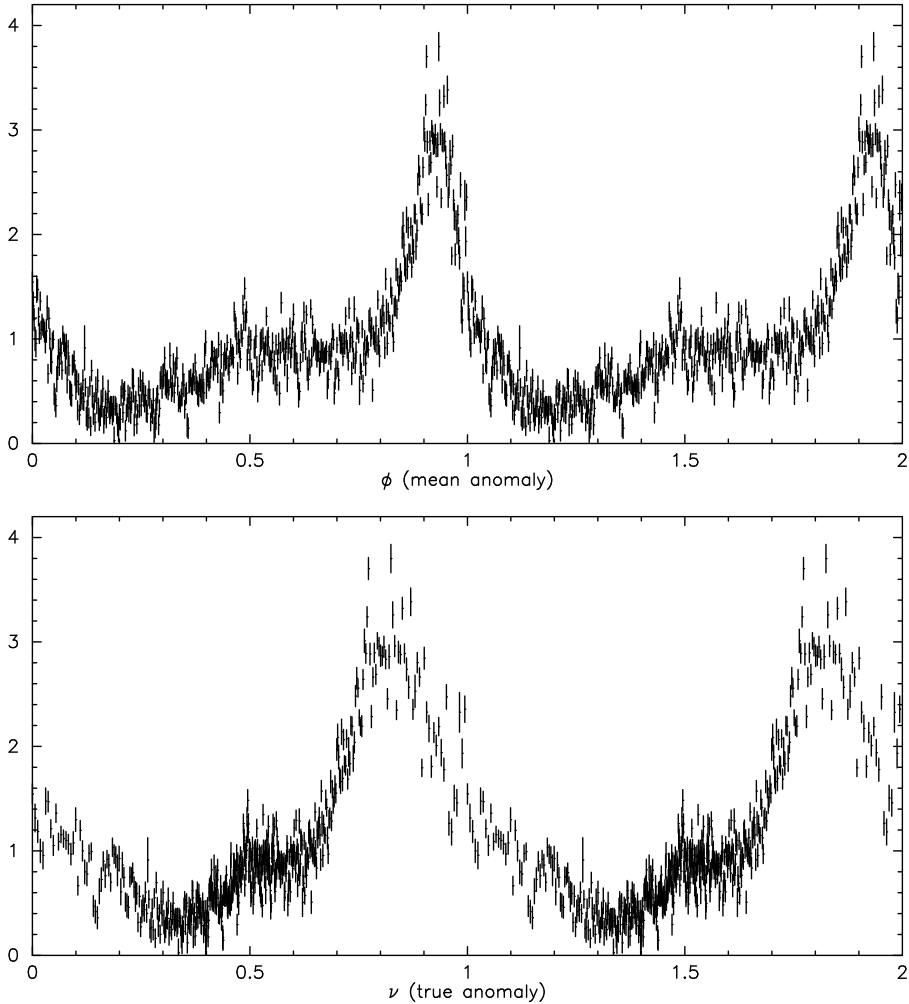}
\caption[]{RXTE/ASM X-ray lightcurve of \object{GX301$-$2} folded with
          the orbital period of the system, as a function of mean
          anomaly $\phi$ ({\it top}) and true anomaly $\nu$ ({\it
          bottom}). The data included in the plot are collected over a
          period of about 9 year (1996--2005) in the energy range
          1.5--12~keV. The peak near periastron passage follows
          naturally from the eccentricity of the orbit, given a
          predicted Bondi-Hoyle X-ray luminosity based on stellar wind
          accretion, but this simple model does not
          reproduce the lightcurve in between the peaks. The reduction
          in X-ray flux following periastron passage ($0.17 \la \phi
          \la 0.33$) is likely due to the absorption of X-rays by the
          dense stellar wind of \wray.}
\label{fig_Xray}
\end{center}
\end{figure*}

\section{Interaction between the dense stellar wind of \wray\ and \object{GX301$-$2}}
\label{interaction}

The X-ray lightcurve of \object{GX301$-$2} includes a strong flare about 1-2 days
before periastron passage; Fig.~\ref{fig_Xray} shows the X-ray
lightcurve of \object{GX301$-$2} obtained with the All Sky Monitor onboard the
{\it Rossi X-ray Timing Explorer} (RXTE). When folding the X-ray
lightcurve with the orbital period of the system, a second, though much
weaker flare appears to be present near apastron passage; this
near-apastron flare is not always detectable \citep{Pravdo01}. Several
stellar wind accretion models have been proposed to explain the
magnitude of the flares and their orbital phase dependence
\citep{Leahy91,Haberl91}. The large eccentricity of the orbit and
Bondi-Hoyle accretion naturally lead to a higher X-ray flux at
periastron passage; an azimuthal velocity component is sufficient to
shift the phase of the main flare \citep{Leahy02}. 

It is, however, not straightforward to explain the increase in X-ray
flux at apastron passage. A dense, slowly expanding disc around \wray\
was postulated by \citet{Pravdo95} to produce an increase in X-ray
flux {\it twice} per orbit, i.e. when the neutron star passes through
the disc, similar to what is observed in some Be/X-ray binaries. It
turns out that the wind plus disc model neither provides an acceptable
fit to the X-ray lightcurve \citep{Pravdo01,Leahy02}. When a gas
stream is introduced in the system, e.g. resulting from tidal
interaction \citep{Layton98}, a better fit is obtained
\citep{Leahy02}. The gas stream flows out from the point on the
primary facing the neutron star \citep{Stevens88} and has a geometry
such that it is crossed twice by the neutron star per orbit,
consistent with a dynamical stream calculation \citep{Leahy02}.  

The X-ray spectrum of \object{GX301$-$2} also provides evidence for the presence
of a gas stream in the system \citep{Saraswat96}. Part of the
low-energy excess below 4~keV detected by the {\it Advanced Satellite
for Cosmology and Astrophysics} (ASCA) can be modelled as due to
scattering of X-rays in the dense stellar wind of \wray. The ultrasoft
emission, however, requires an additional component in the system and
is fitted well by thermal emission from a hot plasma with $kT \sim
0.8$~keV. This emission may arise from X-ray induced shocks in the gas
trailing the neutron star.

\subsection{Orbital modulation of spectral lines formed in the stellar wind}

The observations described above indicate the presence of a gas stream
in the system. If the material in the gas stream is dense and extended
enough, it should produce a signature in spectral lines formed in the
stellar wind of \wray. In May 1996 we monitored the $\ion{He}{i}$
5876~\AA\ and $\rm H \alpha$ line of \wray\ during a periastron passage of
the system, simultaneously with RXTE observations
\citep{Mukherjee04}. In 2002 we covered the optical spectrum of \wray\
with VLT/UVES during one full orbit of the system.
\begin{figure*}[!t]
\begin{center}
\includegraphics[width=6.3cm]{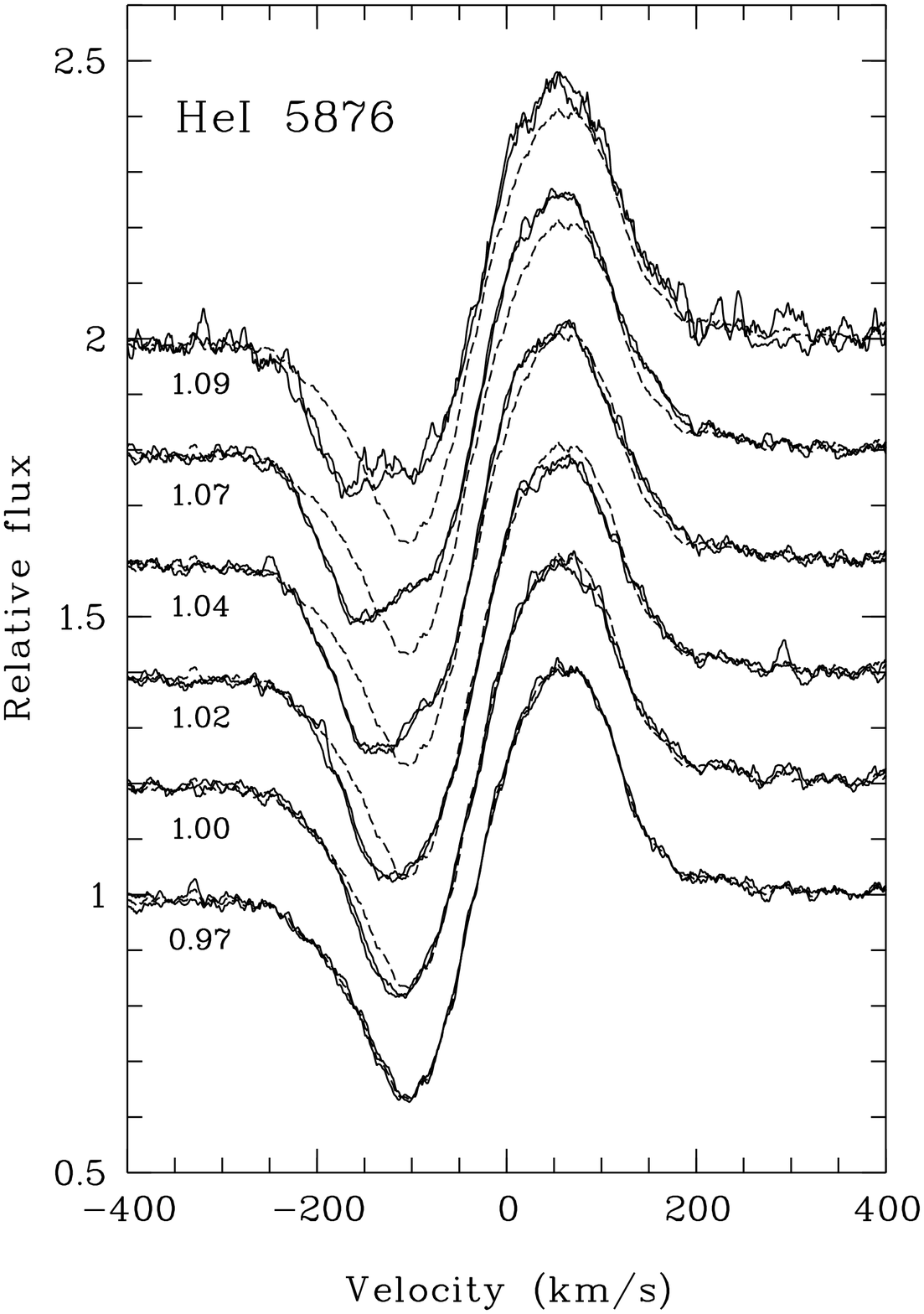}
\includegraphics[width=6.3cm]{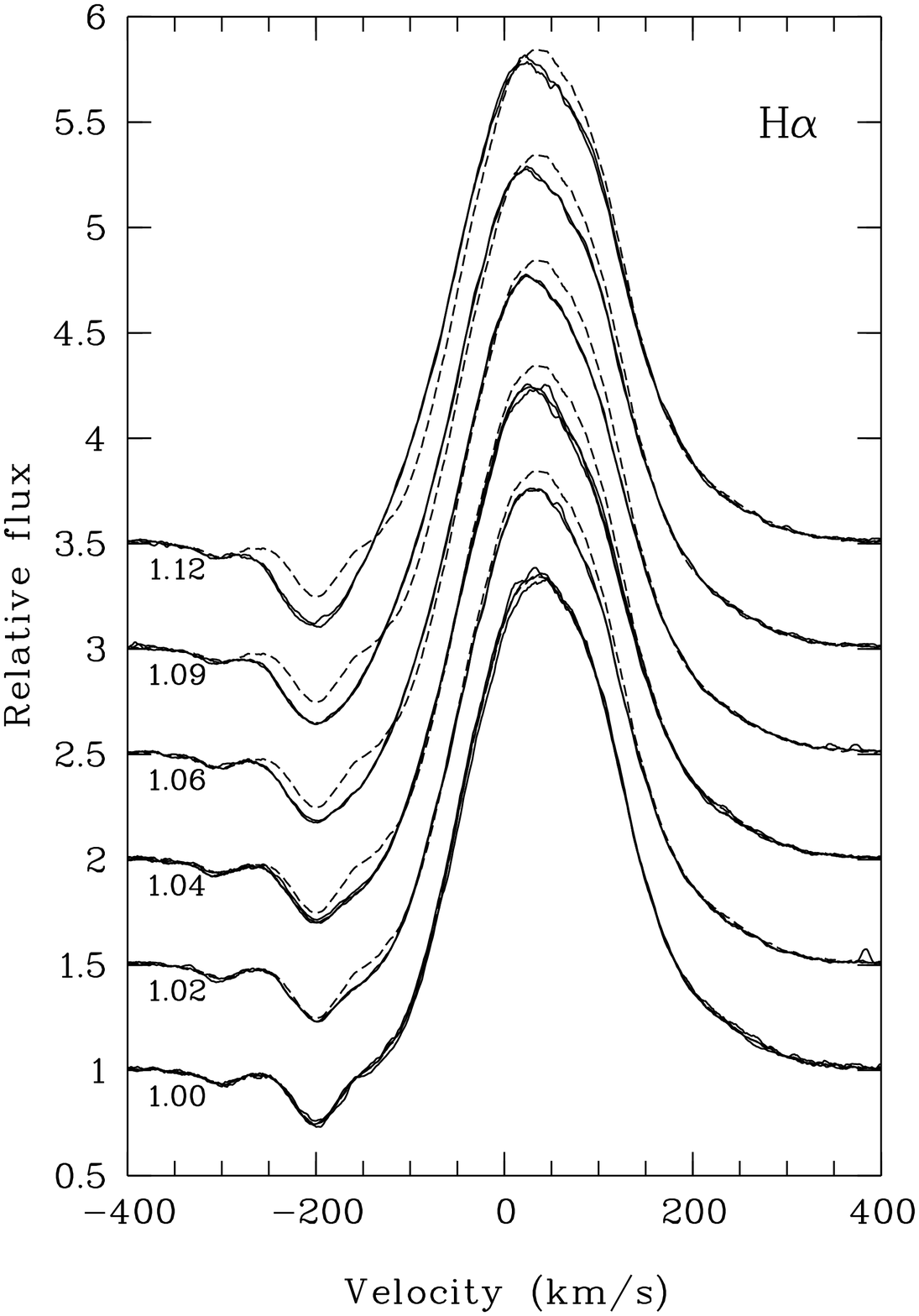}
\caption[]{{\it Left:} Time series of the $\ion{He}{i}$ 5876~\AA\ line
          of \wray\ (\object{GX301$-$2}) obtained in May 1996 with
          CAT/CES. Orbital phase $\phi = 1.0$ corresponds to
          periastron passage. The line of sight to \wray\ is crossed
          by the X-ray source around phase 0.95 (see
          Fig.~\ref{figorbit}). The dashed line represents the
          spectrum at $\phi=0.97$ displayed at the bottom of each
          figure. An additional blue-shifted absorption component
          appears in the absorption trough of the P-Cygni profile at
          $v \sim -150$~\kms\ around $\phi \sim 1$ and moves to higher
          radial velocity with increasing orbital phase. {\it Right:}
          The $\rm H \alpha$ line shows a similar trend.}
\label{figmay96}
\end{center}
\end{figure*}

\subsubsection{Periastron passage in May 1996}

\citet{Mukherjee04} report on archival RXTE observations of
\object{GX301$-$2} using the large-area proportional counter array
(PCA), obtained in the period May 10 to June 15, 1996. Though the
total X-ray luminosity is much higher in the pre-periastron phase,
they do not detect a remarkable change in the intrinsic continuum
spectrum. A considerable increase in column density occurs near
periastron passage, from $N_{\rm H} \sim 10^{23}$~cm$^{-2}$ at $\phi
\sim 0.8$ to $2 \times 10^{24}$~cm$^{-2}$ at $\phi \sim 0.95$. Similar
high values of $N_{\rm H}$ are measured at other orbital phases; e.g. near
$\phi \sim 0.3$, i.e. when the X-ray source is ``behind''
\wray. \citet{Mukherjee04} interpret the large scatter c.q.\ variation
in $N_{\rm H}$ as being due to the clumpiness of the stellar
wind. Additional clumping of the wind near periastron passage may be
due to the extra ionisation of the wind by the X-ray source. The
optical spectrum of \wray\ does, however, not show evidence for
clumping (Sect.~\ref{stellarparameters}).

We obtained high-resolution spectra of the wind lines $\ion{He}{i}$
5876~\AA\ and $\rm H \alpha$ of \wray\ from May 10--16, 1996, i.e. during
the periastron passage covered by
\citet{Mukherjee04}. Fig.~\ref{figmay96} displays a time series of
these two lines; for comparison, the average of the first two spectra
obtained around $\phi=0.97$ is shown as a dashed line. An additional
blue-shifted absorption component (central velocity $v \sim 150$~\kms)
appears near periastron passage ($\phi = 1.0$) and moves to higher
velocity with increasing orbital phase. A similar trend is observed in
$\rm H \alpha$. If the orbital modulation of these wind lines is due to a
gas stream in the system, this stream must cover a significant
fraction of the surface of \wray\ and trail the X-ray source in its
orbit, see Fig.~\ref{figorbit}.

\subsubsection{Monitoring of a full orbit in 2002}

\begin{figure*}[!t]
\begin{center}
\includegraphics[width=6.3cm]{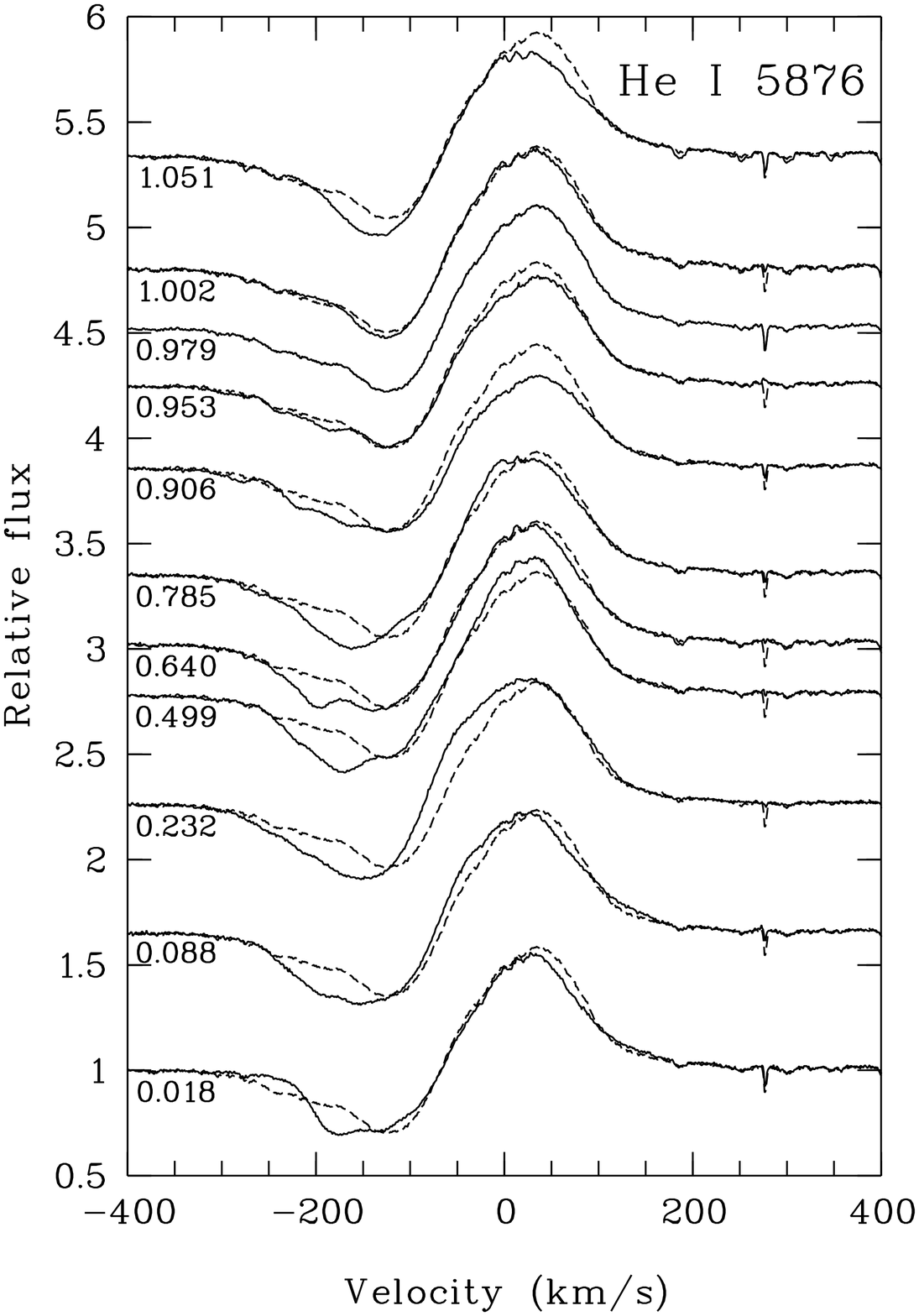}
\includegraphics[width=6.3cm]{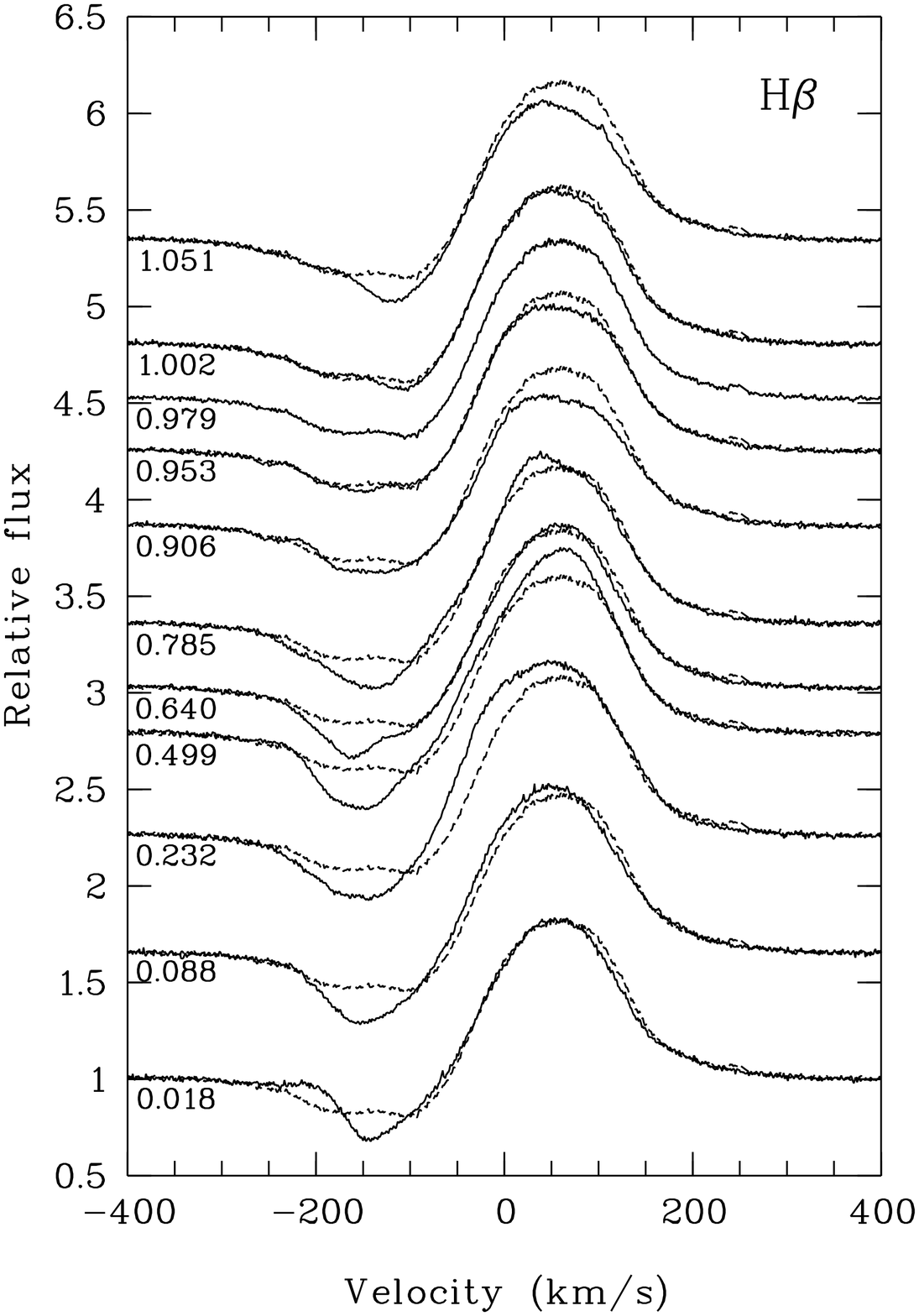}
\caption[]{$\ion{He}{i}$ 5876~\AA\ ({\it left}) and $\rm H \beta$
          ({\it right}) variations over one full orbit of the system
          in the period January - February 2002. The dashed line
          represents the spectrum obtained at orbital phase $\phi =
          0.98$, like in Fig.~\ref{figmay96}. The two lines show an
          almost identical variability pattern. Like in May 1996, the
          blue-shifted absorption increases in strength near
          periastron passage, but remains visible throughout the orbit
          and weakens around $\phi \sim 0.9$. Also note the variations
          in P~Cygni emission: At $\phi=0.23$ additional blue-shifted
          emission is detected, around $\phi=0$ red-shifted emission
          and near $\phi=0.91$ a significant reduction in emission.}
\label{figuves}
\end{center}
\end{figure*}

One would like to know the extent of this gas stream in the system; as
the sightline to \wray\ probes through the system during one
revolution, this question can be addressed. Our VLT/UVES spectra cover
one full orbit of the system in the period January -- February
2002. Fig.~\ref{figuves} displays a time series of $\ion{He}{i}$
5876~\AA\ and $\rm H \beta$; also here the spectrum obtained at $\phi =
0.979$ is shown as a dashed line for comparison. The P~Cygni profiles
of $\ion{He}{i}$ 5876~\AA\ and $\rm H \beta$ are of comparable strength and
show similar variations as observed in May 1996 (Fig.~\ref{figmay96}).
However, the increase in blue-shifted absorption (with respect to the
$\phi=0.979$ spectrum) is also seen at other orbital phases, though
with different strength and velocity structure. This suggests that the
gas stream in the system has a large extent, both in the orbital plane
and in the direction perpendicular to it. The disappearance of the
blue-shifted absorption feature near periastron passage may be due to
the reduction in optical depth of the material contained in the gas
stream when it expands and moves away from the star and, on the other
hand, the relatively small extent of the gas stream near periastron
where it leaves the star.

Variations also occur in the emission part of the profile. Contrary to
the changes observed in the blue-shifted absorption part of the
profile, the material causing these variations is not limited to the
line of sight to \wray. Between $\phi \sim 0.1$ and $\phi \sim 0.25$
additional blue-shifted emission is observed (Fig.~\ref{figuves});
around $\phi \sim 0.9$ an overall reduction in emission is
detected. Near periastron passage the emission has gained in strength
again.  Qualitatively, the variations in line emission are consistent
with the variations observed in absorption and support the conclusion
that an extended structure (gas stream) is present in the system that
trails the neutron star in its orbit (Fig.~\ref{figorbit}). The
observations do not reveal a clue regarding the observed peaks in
X-ray luminosity near peri- and apastron passage, i.e. that these
peaks would be due to the passing of the neutron star through a gas
stream at these locations in the orbit \citep{Leahy02}. On the other
hand, the observations do not exclude this scenario either and
provide the first direct indication for the presence of a gas stream
in the system.

\subsection{X-ray lightcurve and X-ray pulse period history of \object{GX301$-$2}}

\begin{figure}[!t]
\begin{center}
\includegraphics[width=8cm]{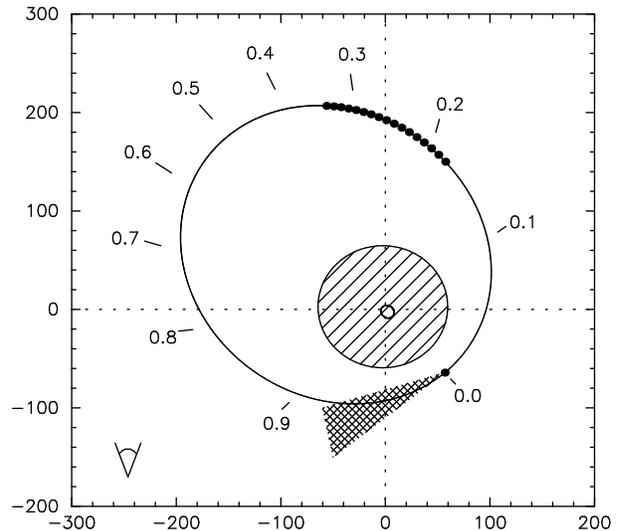}
\caption[]{The eccentric orbit of the X-ray pulsar \object{GX301$-$2}
          around the B hypergiant \wray (shaded). The units along the
          axes are in solar radii; the line of sight to \wray\ is
          indicated by the vertical dotted line. Periastron passage
          occurs at $\phi = 0.0$; in the orbital phase interval $0.18
          \la \phi \la 0.34$ the X-ray source is behind \wray. Part of
          a gas stream is sketched which trails the X-ray source in
          its orbit. The wind line observations suggest that this gas
          stream has a large extent and remains detectable in the line
          of sight to \wray\ during almost one orbital revolution of
          the system.}
\label{figorbit}
\end{center}
\end{figure}

Fig.~\ref{fig_Xray} displays the RXTE/ASM X-ray lightcurve of
\object{GX301$-$2} folded with the orbital period of the system, as a
function of mean and true anomaly. The data points shown are collected
over a period of about 9 years (1996--2005) in the energy range
1.5--12~keV; we refer to \citet{Leahy02} for an earlier version of the
RXTE/ASM lightcurve. The predicted Bondi-Hoyle X-ray luminosity based
on the stellar wind and orbital parameters derived in
Sect.~\ref{modelling} can explain the peak near periastron passage
because of the eccentricity of the orbit, but this simple model does
not reproduce the lightcurve in between the peaks. The reduction in
X-ray flux following periastron passage ($0.17 \la \phi \la 0.33$) is
likely due to the absorption of (soft) X-rays by the dense stellar
wind of \wray\ (Fig.~\ref{figorbit}), especially when one takes into
account the constraints derived on the orbital inclination listed in
Table~\ref{tabmwray}.

Adopting a distance of 3~kpc, the maximum and mean X-ray flux of
\object{GX301$-$2} \citep{Chichkov95} yield a maximum and mean X-ray luminosity
of $L_{\rm max} = 3.9 \times 10^{37}$~erg~s$^{-1}$ and $\langle L_{X}
\rangle = 6.9 \times 10^{36}$~erg~s$^{-1}$. These values are in good
agreement with the predicted X-ray luminosity based on accretion of
the dense hypergiant wind; accretion  from a normal supergiant wind
cannot produce the observed X-ray flux \citep{White85,Kaper98}.
\begin{figure*}[!t]
\begin{center}
\includegraphics[angle=-90,width=10.9cm]{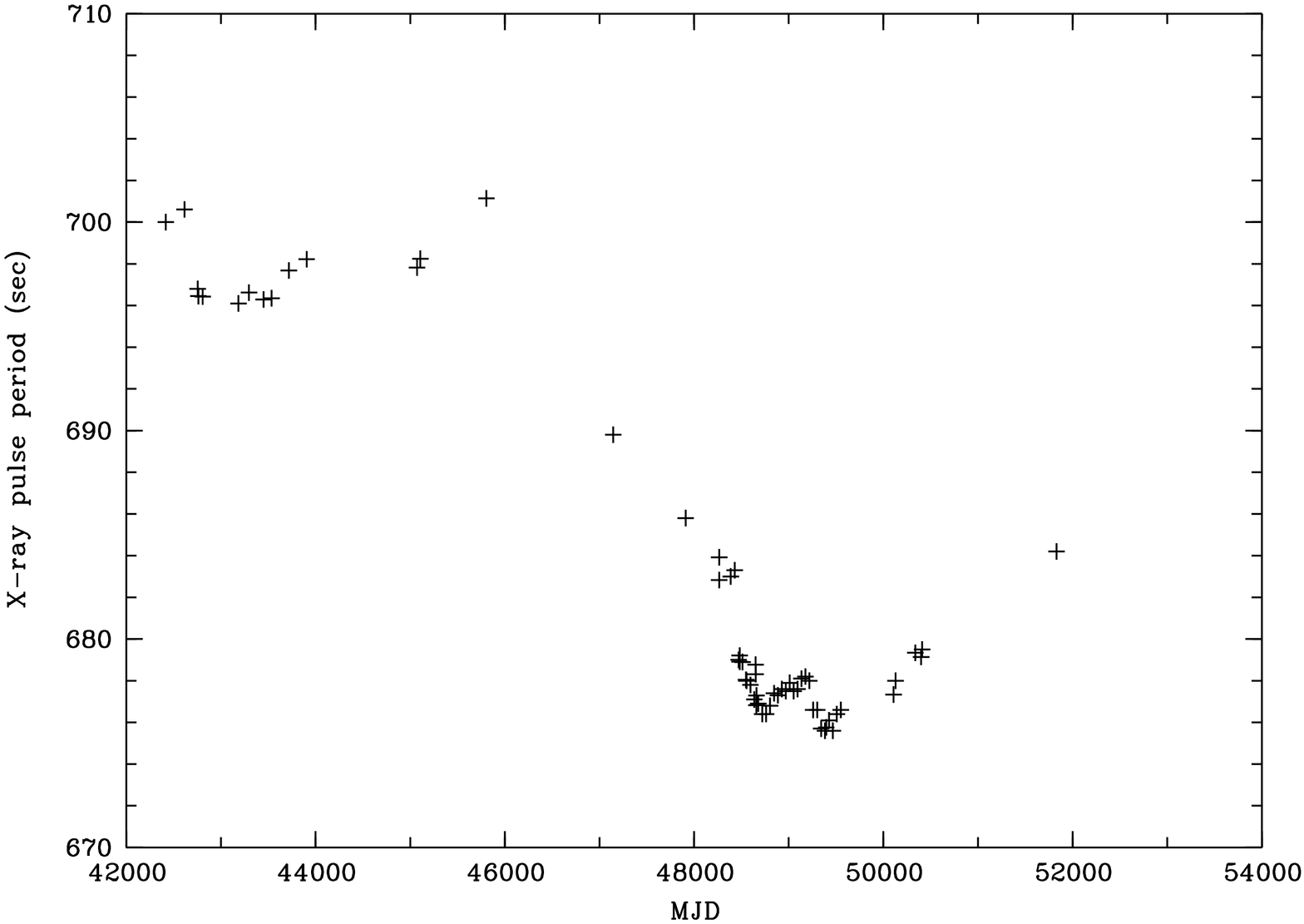}
\\[-0.2cm]
\includegraphics[angle=-90,width=10.9cm]{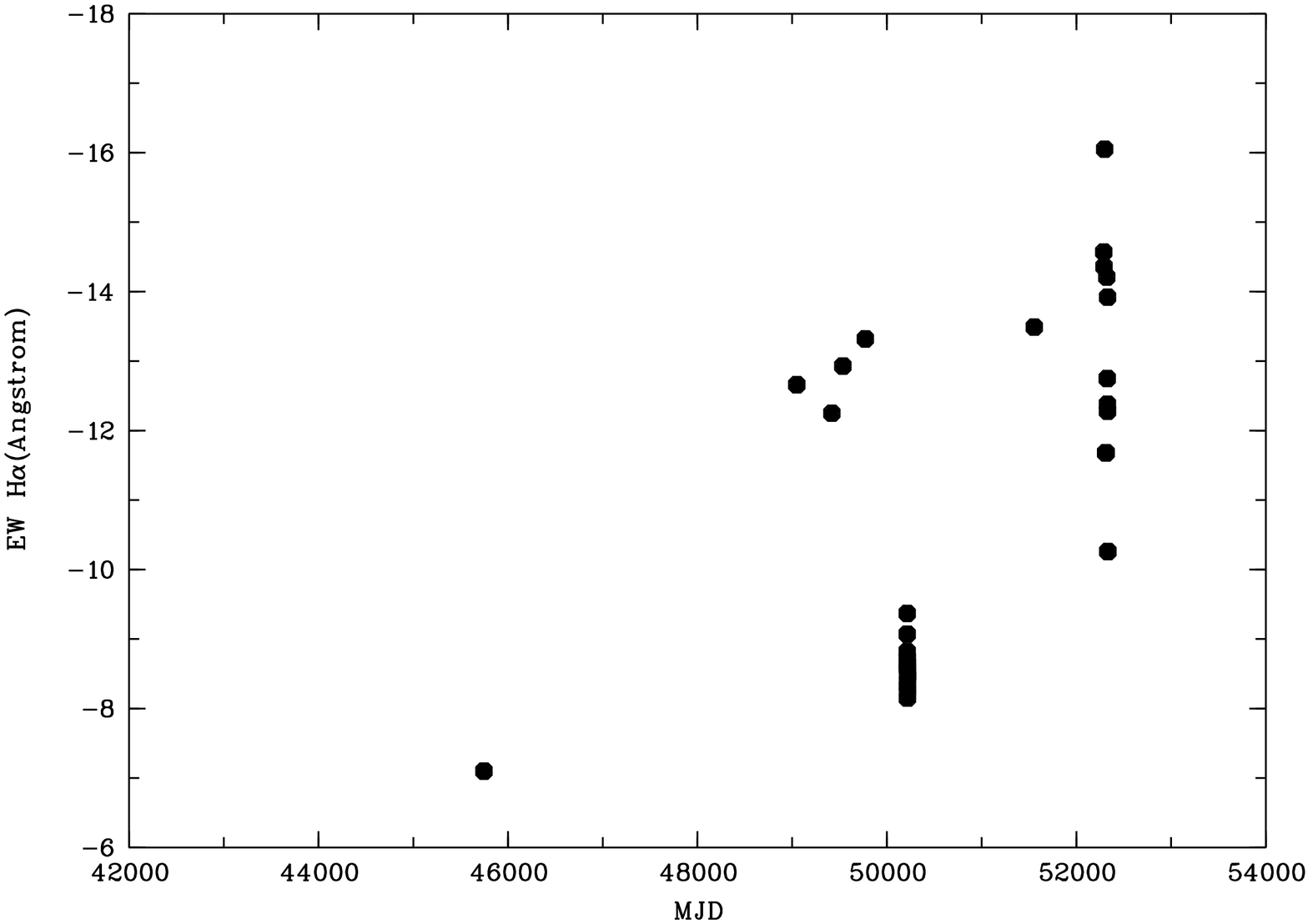}
\caption[]{{\it Top:} The X-ray pulse-period history of
          \object{GX301$-$2} over the period 1975 to 2002. The X-ray
          pulsar spun up from about 700~s in 1985 to 675~s in 1991
          \citep{Koh97}. {\it Bottom:} The $\rm H \alpha$ equivalent
          width has been measured at several short intervals \citep[\ 
          and this paper]{Kaper95}. The $\rm H \alpha$ EW shows
          strong variations and shows an upward trend (i.e. more
          negative and thus stronger in emission), but the time
          sampling is too coarse to derive a correlation.}
\label{figpulse}
\end{center}
\end{figure*}

Not only mass, but also angular momentum is transferred to the X-ray
pulsar. The X-ray pulse period history contains information on the
angular momentum content of the accreting material as a function of
time. Fig.~\ref{figpulse} shows the pulse period of \object{GX301$-$2} as a
function of time \citep{Nagase89,Koh97,Pravdo01,Kreykenbohm04}. In
1984 a long-term spin-up phase started, possibly due to the formation
of an accretion disc. According to \citet{Koh97} the rapid spin-up
episodes measured since 1991, when the X-ray pulsar reached a spin
period of 675~sec, probably represent the formation of transient
accretion discs. The most recent measurements of the X-ray pulse
period suggest that the spin period is increasing again. 

Although \citet{Pravdo01} found some evidence that the average
20--50~keV X-ray intensity may be correlated with the pulse period, it
is interesting to find out whether the long-term trends seen in the
X-ray pulse period history may be due to a change in the mass
accretion rate in this wind-fed system. Fig.~\ref{figpulse} shows the
behaviour of the $\rm H \alpha$ equivalent width (EW) which has been
measured at irregular intervals since 1984 \citep[\ and this
paper]{Kaper95}. $\rm H \alpha$ is a sensitive probe of the wind mass-loss
rate \citep[e.g., \,][]{Puls96}; the more negative the EW, the
stronger the $\rm H \alpha$ emission and the higher the wind mass-loss
rate. The $\rm H \alpha$ EW shows strong variations, even on the relatively
short timescale of the orbit of the system covered by our VLT/UVES
observations. On average, the wind mass-loss rate as reflected by the
$\rm H \alpha$ EW has increased compared to the one observation we have in
February 1984 \citep{Kaper95}. However, the data are not in
contradiction with the simple scenario that the spin-up and spin-down
periods are caused by a change in the overall wind mass-loss rate and
thus a change in the mass and angular momentum accretion rate.

\section{Summary and conclusions}

New optical spectra and infrared photometry strongly support the B
hypergiant classification of \wray\ as proposed by \citet{Kaper95},
even though the radius, distance and luminosity of \wray\ are somewhat
reduced compared to the values listed in that paper: $R = 62 \,
R_{\sun}$, $d \sim 3$~kpc and $L = 5 \times 10^{5} L_{\sun}$. The
radial-velocity curve, though hampered by the intrinsic scatter of
individual measurements, clearly shows the orbital motion of \wray\
from which the radial-velocity amplitude is derived. Combined with the
accurately determined orbital parameters of the X-ray pulsar, the mass
ratio is set to $q = 0.046 \pm 0.014$. The parameter missing to
calculate the masses of the two stars is the orbital inclination. The
system is not eclipsing, though the X-ray light curve shows a
reduction in X-ray flux during the orbital phase interval when the
X-ray source is behind the B hypergiant. That the orbital inclination
indeed cannot be very low follows from the argument that the mass of
the neutron star must be less than 3.2~M$_{\sun}$ (causality limit)
and likely even less than 2.5~M$_{\sun}$ (maximum neutron star mass
based on equation of state). As a consequence, the mass of \wray\ is
less than 68 (53)~M$_{\sun}$ and higher than 39~M$_{\sun}$ (no X-ray
eclipse). The lower limit on the mass of the neutron star is $1.85 \pm
0.6$~M$_{\sun}$, suggesting that the neutron star belongs to the
high-mass peak in the bimodal neutron-star mass distribution as
proposed by \citet{Timmes96}, like Vela~X-1 \citep{Barziv01} and
probably 4U1700-37 \citep{Clark02}.

\citet{VandenHeuvel84} propose that the system evolved from a 42+38
M$_{\sun}$ binary, the lowest-mass realistic progenitor system
fulfilling the condition that tandem evolution is avoided in a
case~B scenario. \citet{Wellstein99} propose that the system evolved
in a case~A scenario, leading to a progenitor system of initially
25+24 M$_{\sun}$. The phase of mass transfer has to be fully
conservative in order to reproduce the current mass of \wray\ and its
position in the Hertzsprung-Russell diagram. The surface chemical
abundances of \wray\ derived from our spectra agree with the
predictions of \citet{Wellstein99}. Therefore, the progenitor mass of
the neutron star may indeed be as low as 25~M$_{\sun}$, rather than
the 50~M$_{\sun}$ proposed by \citet{Kaper95}. As a consequence, the
lower limit for black-hole formation in a massive binary, derived from
this system, becomes 25~M$_{\sun}$.

The orbital modulation of spectral lines formed in the stellar wind of
\wray, such as $\rm H \beta$ and $\ion{He}{i}$ 5876~\AA, indicate the presence
of a gas stream in the system. Such a gas stream was proposed to
explain the peaks in the X-ray light curve near apastron and
periastron. We find no observational evidence for an extended
equatorial disc surrounding \wray. The B hypergiant is not in
corotation with the X-ray pulsar during periastron passage, which may
be a complicating factor for models explaining the presence of a gas
stream in the system originating from tidal interaction
\citep{Layton98}. However, it may well be that \wray\ exceeds its
tidal lobe during periastron passage, enabling the formation of a gas
stream in the system. The decrease in X-ray pulse period of \object{GX301$-$2}
from 1984 to 1991 may be due to an increase in the mass and angular
momentum accretion rate related to a higher wind mass-loss rate of
\wray.

\begin{acknowledgements}
We thank Paul Crowther and Mike Barlow for their help in interpreting
the interstellar spectrum in this line of sight. Marten van Kerkwijk
and Ed van den Heuvel are acknowledged for helpful discussions. We
thank John Hillier for providing his atmospheric code. We thank an
anonymous referee for carefully reading the manuscript. LK has
been supported by a fellowship from the Royal Academy of Arts and
Sciences. AvdM acknowledges financial support from the Dutch Research
School for Astronomy (NOVA). FN received support from
PNAYA2003-02785-E and AYA2004-08271-C02-02 and the Ramon y Cajal program.
\end{acknowledgements}

\bibliography{./references_lex}

\end{document}